\documentclass[12pt]{article}

\def\ltwid{\mathrel{\raise.3ex\hbox{$<$\kern-.75em\lower1ex\hbox{$\sim$}}}}
\def\gtwid{\mathrel{\raise.3ex\hbox{$>$\kern-.75em\lower1ex\hbox{$\sim$}}}}

\def\square{\kern1pt\vbox{\hrule height 1.2pt\hbox{\vrule width 1.2pt\hskip 3pt
   \vbox{\vskip 6pt}\hskip 3pt\vrule width 0.6pt}\hrule height 0.6pt}\kern1pt}

\begin{document}

\begin{titlepage}

\begin{flushright}
ITP-UU-12/02, SPIN-12/02 \\
UFIFT-QG-11-05
\end{flushright}

\begin{center}
{\bf The Coincidence Limit of the Graviton Propagator in de Donder
Gauge on de Sitter Background}
\end{center}

\begin{center}
E. O. Kahya$^*$
\end{center}

\begin{center}
Theoretisch-Physikalisches Institute,
Friedrich-Schiller-Universit\"at Jena, \\
Max-Wien-Platz 1, D-07743 Jena, GERMANY
\end{center}

\begin{center}
S. P. Miao$^{\dagger}$
\end{center}

\begin{center}
\it{Institute for Theoretical Physics \& Spinoza Institute, Utrecht
University \\ Leuvenlaan 4, Postbus 80.195, 3508 TD Utrecht, NETHERLANDS}\\
\end{center}

\begin{center}
R. P. Woodard$^{\ddagger}$
\end{center}

\begin{center}
\it{Department of Physics, University of Florida \\
Gainesville, FL 32611, UNITED STATES}
\end{center}

\begin{center}
ABSTRACT
\end{center}
We explicitly work out the de Sitter breaking contributions to the
recent solution for the de Donder gauge graviton propagator on de
Sitter. We also provide explicit power series expansions for the two
structure functions, which are suitable for implementing dimensional
regularization. And we evaluate the coincidence limit of the
propagator.

\begin{flushleft}
\end{flushleft}

\begin{flushleft}
$^*$ e-mail: emre-onur.kahya@uni-jena.de \\
$^{\dagger}$ e-mail: S.Miao@uu.nl \\
$^{\ddagger}$ e-mail: woodard@phys.ufl.edu
\end{flushleft}

\end{titlepage}

\section{Introduction}

The naive Fourier mode sum for the propagator of a tachyonic scalar
field fails to converge in flat space. No one considers this
surprising because the modes with $\Vert \vec{k} \Vert^2 < -M^2$ are
unstable. What these degrees of freedom really do is classically
roll down the negative potential. The propagator equation can be
solved, but the solutions break Poincar\'e invariance.

This sort of instability is more pronounced in an expanding universe
with scale factor $a(t)$ because the physical momentum is $\Vert
\vec{k} \Vert/a(t)$, so all modes are eventually redshifted past
stability. One symptom of the problem is that even the massless,
minimally coupled scalar suffers from it for sufficiently small
deceleration \cite{FP}. As for flat space tachyons, the propagator
equation can be solved but the solutions involve time dependent
terms associated with the instability \cite{JMPW}.

An interesting aspect of the problem is that, except for certain
discrete values of the deceleration parameter, the use of analytic
continuation techniques gives a formal solution to the propagator
equation without this time dependence \cite{Tag}. In fact the naive
Fourier mode sum is always infrared divergent but, for most values
of the deceleration parameter this divergence is of the power law
type which drops out when analytic continuation techniques are
employed. The problematic discrete values of the deceleration
parameter are just those for which either the primary infrared
divergence, or a subdominant one, happens to be logarithmic
\cite{JMPW}. So the correct result is that the formal solutions
which continue around power law divergences do not represent true
propagators, which is of course liable to happen whenever one only
solves the propagator equation without constructing it from a mode
sum \cite{TW1}.

The de Sitter geometry brings the problem into sharp focus because
it is the most negative deceleration consistent with classical
stability and because it possesses a maximal isometry group
analogous to Poincar\'e invariance. The breaking of de Sitter
invariance was first noted in the coincidence limit of the massless,
minimally coupled scalar propagator in 1982 \cite{first}. Five years
later Allen and Folacci gave a formal proof that no de Sitter
invariant solution exists \cite{AF}. As with sufficiently small
deceleration, endowing the scalar with a tachyonic mass gives formal
de Sitter invariant solutions, except at the discrete values of the
mass for which one of the power law infrared divergences becomes
logarithmic \cite{MTW1}. Also as before, these formal solutions are
not true propagators.

Because free dynamical gravitons obey the same equation as massless,
minimally coupled scalars \cite{Grishchuk} it has long been obvious
that the graviton propagator must suffer from the same problem as
its scalar cousin \cite{RPW}. Indeed, an explicit mode sum
construction for the graviton propagator in a non-de Sitter
invariant gauge \cite{TW2} shows physical de Sitter breaking even
after the compensating gauge transformation is added to restore the
gauge condition \cite{Kleppe}. This conclusion was for years
disputed by mathematical physicists on the grounds that they could
find explicit, de Sitter invariant solutions by adding covariant 
gauge fixing terms to the action and then analytically continuing 
from Euclidean space \cite{INVPROP}. However, it was recently 
demonstrated that there is an obstacle to adding covariant gauge 
fixing terms to a gauge theory on any manifold which suffers from 
a linearization instability \cite{MTW2}. Ignoring this problem in 
Feynman gauge \cite{AJ} leads to unphysical singularities at one 
loop order in scalar quantum electrodynamics \cite{KW1}.

One can still enforce exact gauge conditions, but it was long ago
found that insisting on a de Sitter invariant solution for certain
exact gauges results in infrared divergences \cite{IAEM,Folacci}.
That conclusion was also dismissed by mathematical physicists on the
grounds that the problem is limited to only discrete values of the
two parameter family of covariant gauge conditions \cite{Higuchi}.
However, we can now recognize that, like the case of tachyonic
scalars, the naive mode sum is always infrared divergent --- and
hence invalid. The only distinction of the special values of the
gauge parameters is that, for these values one of the power law
divergences happens to become logarithmic \cite{MTW1}. So the
correct conclusion in all cases is that the graviton propagator
breaks de Sitter invariance, and this was recently demonstrated by
an explicit solution in de Donder gauge \cite{MTW3}.

The goal of this paper is to put the de Donder gauge propagator
\cite{MTW3} into a tractable form from which dimensional
regularization computations can be performed. We shall also make the
tensor structure of the de Sitter breaking parts explicit, and we
shall evaluate the coincidence limit of the full propagator. The
final results seem quite a bit more complicated than when a simple
de Sitter breaking gauge is employed \cite{TW2}. Two motivations for
developing this gauge are:
\begin{itemize}
\item{Avoiding the noninvariant counterterms which occur with a de
Sitter breaking gauge; and}
\item{Being able to check existing results \cite{TW3,TW4,MW1,KW2}
for gauge dependence.}
\end{itemize}
It might also be that the apparent complications of this gauge drop
out when all the derivatives are acted and the indices are
contracted. Exactly that occurs when using a de Sitter invariant
solution for the Lorentz gauge photon propagator \cite{TW5} to
perform one and even two loop computations in scalar quantum
electrodynamics \cite{PTW}.

This paper has seven sections of which the first is this
introduction. In section 2 we review notation and some previous
results \cite{MTW1,MTW2,MTW3} of great relevance to the current
work. We also explain how the graviton propagator can be written in
terms of differential projectors acting on a spin zero and a spin
two structure function. Section 3 derives explicit results for each
of the two structure functions. In sections 4 and 5 we act the
differential projectors on the de Sitter invariant and de Sitter
breaking parts. (Many technical details of this analysis are
consigned to an Appendix.) The coincidence limit is taken in section
6, and section 7 gives our discussion.

\section{Notation}

This section reviews and consolidates notation and results
introduced in earlier work. We begin by describing the coordinate
system and tensor basis employed in this paper. We then present the
solution for a general scalar propagator with a possibly tachyonic
mass, and describe how to integrate such propagators. The section
closes with a review of the general form of the graviton propagator
in de Donder gauge.

\subsection{Working on de Sitter}

We work on the $D$-dimensional open conformal submanifold in which
de Sitter can be imagined as a special case of the larger class of
homogeneous, isotropic and spatially flat geometries relevant to
cosmology. A spacetime point $x^{\mu} = (x^0, x^i)$ takes values in
the ranges,
\begin{equation}
-\infty < x^0 < 0 \quad {\rm and} \quad  -\infty < x^i < +\infty
\quad {\rm for} \quad i = 1,\ldots,(D\!-\!1) \; .
\end{equation}
In these coordinates the invariant element is,
\begin{equation}
ds^2 \equiv g_{\mu\nu} dx^{\mu} dx^{\nu} = a_x^2 \Bigl[-(dx^0)^2 +
d\vec{x} \!\cdot\! d\vec{x} \Bigr] = a_x^2 \eta_{\mu\nu} dx^{\mu}
dx^{\nu}\; ,
\end{equation}
where $\eta_{\mu\nu}$ is the Lorentz metric and $a_x \equiv -1/Hx^0$
is the scale factor.

Although infrared divergences do introduce de Sitter breaking into
the graviton propagator they do so in a limited way that leaves the
largest part of the result de Sitter invariant. For dimensional
regularization computations it is best to express this de Sitter
invariant part in terms of the length function $y(x;z)$,
\begin{equation}
y(x;z) \equiv a_x a_z H^2 \Biggl[ \Bigl\Vert \vec{x} \!-\! \vec{z}
\Bigr\Vert^2 - \Bigl(\vert x^0 \!-\! z^0\vert \!-\! i
\epsilon\Bigr)^2 \Biggr]\; . \label{ydef}
\end{equation}
Except for the factor of $i\epsilon$ (whose purpose is to enforce
Feynman boundary conditions) the function $y(x;z)$ is closely
related to the invariant length $\ell(x;z)$ from $x^{\mu}$ to
$z^{\mu}$,
\begin{equation}
y(x;z) = 4 \sin^2\Bigl( \frac12 H \ell(x;z)\Bigr) \; .
\end{equation}

With this de Sitter invariant quantity $y(x;z)$, we can form a
convenient basis of de Sitter invariant bi-tensors. Note that
because $y(x;z)$ is de Sitter invariant, so too are covariant
derivatives of it. With the metrics $g_{\mu\nu}(x)$ and
$g_{\mu\nu}(z)$, the first three derivatives of $y(x;z)$ furnish a
convenient basis of de Sitter invariant bi-tensors \cite{KW1},
\begin{eqnarray}
\frac{\partial y(x;z)}{\partial x^{\mu}} & = & H a_x \Bigl(y
\delta^0_{\mu}
\!+\! 2 a_z H \Delta x_{\mu} \Bigr) \; , \label{dydx} \\
\frac{\partial y(x;z)}{\partial z^{\nu}} & = & H a_z \Bigl(y
\delta^0_{\nu}
\!-\! 2 a_x H \Delta x_{\nu} \Bigr) \; , \label{dydz} \\
\frac{\partial^2 y(x;z)}{\partial x^{\mu} \partial z^{\nu}} & = &
H^2 a_x a_z \Bigl(y \delta^0_{\mu} \delta^0_{\nu} \!+\! 2 a_z H
\Delta x_{\mu} \delta^0_{\nu} \!-\! 2 a_x \delta^0_{\mu} H \Delta
x_{\nu} \!-\! 2 \eta_{\mu\nu}\Bigr) \; . \qquad \label{dydxdx'}
\end{eqnarray}
Here and subsequently we define $\Delta x_{\mu} \equiv \eta_{\mu\nu}
(x \!-\!z)^{\nu}$. Acting covariant derivatives generates more basis
tensors, for example \cite{KW1},
\begin{eqnarray}
\frac{D^2 y(x;z)}{Dx^{\mu} Dx^{\nu}}
& = & H^2 (2 \!-\!y) g_{\mu\nu}(x) \; , \label{covdiv} \\
\frac{D^2 y(x;z)}{Dz^{\mu} Dz^{\nu}} & = & H^2 (2 \!-\!y)
g_{\mu\nu}(z) \; .
\end{eqnarray}
The contraction of any pair of the basis tensors also produces more
basis tensors \cite{KW1},
\begin{eqnarray}
g^{\mu\nu}(x) \frac{\partial y}{\partial x^{\mu}} \frac{\partial
y}{\partial x^ {\nu}} & = & H^2 \Bigl(4 y - y^2\Bigr) =
g^{\mu\nu}(z) \frac{\partial y}{
\partial z^{\mu}} \frac{\partial y}{\partial z^{\nu}} \; ,
\label{contraction1}\\
g^{\mu\nu}(x) \frac{\partial y}{\partial x^{\nu}} \frac{\partial^2
y}{
\partial x^{\mu} \partial z^{\sigma}} & = & H^2 (2-y) \frac{\partial y}{
\partial z^{\sigma}} \; ,
\label{contraction2}\\
g^{\rho\sigma}(z) \frac{\partial y}{\partial z^{\sigma}}
\frac{\partial^2 y}{\partial x^{\mu} \partial z^{\rho}} & = & H^2
(2-y) \frac{\partial y}{\partial x^{\mu}} \; ,
\label{contraction3}\\
g^{\mu\nu}(x) \frac{\partial^2 y}{\partial x^{\mu} \partial
z^{\rho}} \frac{\partial^2 y}{\partial x^{\nu} \partial z^{\sigma}}
& = & 4 H^4 g_{\rho\sigma}(z) - H^2 \frac{\partial y}{\partial
z^{\rho}} \frac{\partial y}{\partial z^{\sigma}} \; ,
\label{contraction4}\\
g^{\rho\sigma}(z) \frac{\partial^2 y}{\partial x^{\mu}\partial
z^{\rho}} \frac{\partial^2 y}{\partial x^{\nu} \partial z^{\sigma}}
& = & 4 H^4 g_{\mu\nu}(x) - H^2 \frac{\partial y}{\partial x^{\mu}}
\frac{\partial y}{\partial x^{\nu}} \; . \label{contraction5}
\end{eqnarray}

The tensor structure of de Sitter breaking terms requires
derivatives of the quantity $u(x;z) \equiv \ln(a_x a_z)$,
\begin{equation}
\frac{\partial u}{\partial x^{\mu}} = H a_x \delta^0_{\mu} \qquad ,
\qquad \frac{\partial u}{\partial z^{\rho}} = H a_z \delta^0_{\rho}
\; . \label{uders}
\end{equation}
Covariant derivatives of the new tensors involve some extra
identities in addition to those of $y(x;z)$ \cite{MTW1},
\begin{equation}
\frac{D^2 u}{D x^{\mu} D x^{\nu}} = -H^2 g_{\mu\nu}(x) -
\frac{\partial u}{\partial x^{\mu}} \frac{\partial u}{\partial
x^{\nu}} \quad , \quad \frac{D^2 u}{D z^{\mu} D z^{\nu}} = -H^2
g_{\mu\nu}(z) - \frac{\partial u}{\partial z^{\mu}} \frac{\partial
u}{\partial z^{\nu}} \; .
\end{equation}
There are also some new contraction identities,
\begin{eqnarray}
g^{\mu\nu}(x) \frac{\partial u}{\partial x^{\mu}} \frac{\partial
u}{\partial x^{\nu}} & = & - H^2 = g^{\rho\sigma}(z) \frac{\partial
u}{\partial z^{\rho}} \frac{\partial u}{\partial z^{\sigma}}
\; , \\
g^{\mu\nu}(x) \frac{\partial u}{\partial x^{\mu}} \frac{\partial
y}{\partial x^{\nu}} & = & - H^2 \Bigl[ y \!-\! 2
+ 2 \frac{a_z}{a_x} \Bigr] \; , \\
g^{\rho\sigma}(z) \frac{\partial u}{\partial z^{\rho}}
\frac{\partial y}{\partial z^{\sigma}} & = & - H^2 \Bigl[ y \!-\! 2
+ 2 \frac{a_x}{a_z} \Bigr] \; , \\
g^{\mu\nu}(x) \frac{\partial u}{\partial x^{\mu}} \frac{\partial^2
y}{\partial x^{\nu} \partial z^{\rho}} & = & - H^2 \Bigl[
\frac{\partial y}{\partial z^{\rho}} + 2 \frac{a_z}{a_x}
\frac{\partial u}{\partial z^{\rho}} \Bigr] \; , \\
g^{\rho\sigma}(z) \frac{\partial u}{\partial z^{\rho}}
\frac{\partial^2 y}{\partial x^{\mu} \partial z^{\sigma}} & = & -
H^2 \Bigl[ \frac{\partial y}{\partial x^{\mu}} + 2 \frac{a_x}{a_z}
\frac{\partial u}{\partial x^{\mu}} \Bigr] \; .
\end{eqnarray}

\subsection{General Scalar Propagators}

We work with a general scalar propagator $i\Delta_b(x;z)$ which
obeys the equation,
\begin{equation}
\Bigl[ \square + (b^2 \!-\! b_A^2) H^2 \Bigr] i\Delta_b(x;z) =
\frac{i \delta^D(x \!-\! z)}{\sqrt{-g}} \; .
\end{equation}
Here and henceforth we define the index $b_A \equiv \frac{D-1}2$.
For the case of $b < b_A$ the propagator has a positive mass-squared
and its propagator is de Sitter invariant. We display its expansion
for $b = \nu$,
\begin{eqnarray}
\lefteqn{i\Delta^{\rm dS}_{\nu}(x;z) = \frac{H^{D-2}}{(4
\pi)^{\frac{D}2}} \Biggl\{ \Gamma\Bigl(\frac{D}2 \!-\! 1\Bigr)
\Bigl(\frac{4}{y}\Bigr)^{\frac{D}2-1} - \frac{\Gamma(\frac{D}2)
\Gamma(1 \!-\! \frac{D}2)}{ \Gamma(\frac12 \!+\! \nu) \Gamma(\frac12
\!-\! \nu)} \sum_{n=0}^{\infty} } \nonumber \\
& & \hspace{-.7cm} \times \Biggl[ \frac{\Gamma(\frac32 \!+\! \nu
\!+\! n) \Gamma(\frac32 \!-\! \nu \!+\! n)}{ \Gamma(3 \!-\!
\frac{D}2 \!+\! n) \, (n \!+\! 1)!} \Bigl(\frac{y}4 \Bigr)^{n -
\frac{D}2 +2} \!\!\!-\! \frac{\Gamma(b_A \!+\! \nu \!+\! n)
\Gamma(b_A \!-\! \nu \!+\! n)}{\Gamma(\frac{D}2 \!+\! n) \, n!}
\Bigl(\frac{y}4\Bigr)^n \Biggr] \! \Biggr\} . \qquad
\label{expansion}
\end{eqnarray}
For $b \geq b_A$ the naive mode sum would be infrared divergent so
one must add a de Sitter breaking, infrared correction
\cite{AF,MTW1}. We have constructed this correction to preserve the
symmetries of homogeneity and isotropy,
\begin{eqnarray}
\lefteqn{\Delta^{\rm IR}_{\nu}(x;z) =
\frac{H^{D-2}}{(4\pi)^{\frac{D}2}} \frac{\Gamma(\nu)
\Gamma(2\nu)}{\Gamma(b_A)
\Gamma(\nu \!+\! \frac12)} \times \theta(\nu \!-\! b_A) } \nonumber \\
& & \hspace{.7cm} \times \sum_{N=0}^{[\nu - b_A]} \frac{(a_x
a_z)^{\nu - b_A - N}}{\nu \!-\! b_A \!-\! N} \sum_{n=0}^N \Bigl(
\frac{a_x}{a_z} \!+\! \frac{a_z}{a_x}\Bigr)^n \sum_{m=0}^{
[\frac{N-n}2]} C_{Nnm} (y \!-\!2)^{N-n-2m} \; , \qquad
\label{series}
\end{eqnarray}
where the coefficients $C_{Nnm}$ are,
\begin{eqnarray}
\lefteqn{C_{Nnm} = \frac{(-\frac14)^N}{m! n! (N \!-\!n \!-\! 2m)!}
\times \frac{\Gamma(b_A \!+\! N \!+\! n \!-\!
\nu)}{\Gamma(b_A \!+\! N \!-\! \nu)} } \nonumber \\
& & \hspace{2cm} \times \frac{\Gamma(b_A)}{\Gamma(b_A \!+\! N\!-\!
2m)} \times \frac{\Gamma(1 \!-\!\nu)}{\Gamma(1 \!-\! \nu \!+\! n
\!+\! 2m)} \times \frac{\Gamma(1 \!-\! \nu)}{\Gamma(1 \!-\! \nu
\!+\! m)} \; . \qquad \label{cdef}
\end{eqnarray}
The full propagator is therefore,
\begin{equation}
i\Delta_b(x;z) = \lim_{\nu \rightarrow b} \Bigl[ i\Delta^{\rm
dS}_{\nu}(x;z) + \Delta^{\rm IR}_{\nu}(x;z) \Bigr] \; .
\end{equation}

\subsection{Integrating Scalar Propagators}

Many of the ``propagators'' we employ are actually integrated
propagators which obey the equation,
\begin{equation}
\Bigl[ \square + (b^2 \!-\! b_A^2) H^2 \Bigr] i \Delta_{bc}(x;z) =
i\Delta_c(x;z) \; . \label{Int1}
\end{equation}
The solution is easily seen to be \cite{MTW1,MTW2},
\begin{equation}
i\Delta_{bc}(x;z) = \frac1{(b^2 \!-\! c^2) H^2} \Bigl[
i\Delta_c(x;z) \!-\! i\Delta_b(x;z)\Bigr] = i\Delta_{cb}(x;z) \; .
\label{Int2}
\end{equation}
For the special case that the indices $b$ and $c$ agree one gets a
derivative,
\begin{equation}
i\Delta_{bb}(x;z) = -\frac1{2 b H^2} \frac{\partial}{\partial b}
i\Delta_b(x;z) \; . \label{Int3}
\end{equation}

We also employ a doubly integrated propagator which obeys the
equation,
\begin{equation}
\Bigl[ \square + (b^2 \!-\! b_A^2) H^2 \Bigr] i \Delta_{bcd}(x;z) =
i\Delta_{cd}(x;z) \; . \label{Int4}
\end{equation}
The solution can be written in a form which is manifestly symmetric
under any interchange of the three indices $a$, $b$ and $c$,
\begin{eqnarray}
i\Delta_{bcd}(x;z) &\!\! = \!\!& \frac{ i\Delta_{bd}(x;z) \!-\!
i\Delta_{bc}(x;z)}{(c^2 \!-\! d^2) H^2} \; , \qquad \label{Int5a} \\
& \!\!=\!\! & \frac{ (d^2 \!-\! c^2) i\Delta_{b}(x;z) \!+\! (b^2
\!-\! d^2) i\Delta_{c}(x;z) \!+\! (c^2 \!-\! b^2) i\Delta_{d}(x;z)
}{(b^2 \!-\! c^2) (c^2 \!-\! d^2) (d^2 \!-\! b^2) H^4} \; . \qquad
\label{Int5b}
\end{eqnarray}
The case in which two of the indices are the same gives,
\begin{equation}
i\Delta_{bcc}(x;z) = -\frac1{2 c H^2} \frac{\partial}{\partial c}
i\Delta_{bc}(x;z) = \frac{ i\Delta_{cc}(x;z) \!-\!
i\Delta_{bc}(x;z)}{ (b^2 \!-\! c^2) H^2} \; . \label{Int6}
\end{equation}
And equating all three indices produces,
\begin{eqnarray}
i\Delta_{bbb}(x;z) & = & -\frac1{2 b H^2} \frac{\partial}{\partial
b} i\Delta_{bc}(x;z) \Bigl\vert_{c = b} \; , \qquad \label{Int7a} \\
& = & -\frac1{8 b^3 H^4} \Biggl[ \frac{\partial}{\partial b}
i\Delta_b(x;z) \!-\! b \Bigl(\frac{\partial}{\partial b}\Bigr)^2
i\Delta_b(x;z) \Biggr] \; . \label{Int7b}
\end{eqnarray}

\subsection{Form of the Graviton Propagator}

The graviton propagator in de Donder gauge can be expressed as the
sum of a spin zero part and a spin two part,
\begin{equation}
i \Bigl[\mbox{}_{\alpha\beta} \Delta_{\rho\sigma} \Bigr](x;z) = i
\Bigl[\mbox{}_{\alpha\beta} \Delta^0_{\rho\sigma} \Bigr](x;z) + i
\Bigl[\mbox{}_{\alpha\beta} \Delta^2_{\rho\sigma} \Bigr](x;z) \; .
\label{gravdecomp}
\end{equation}
Each part is represented as product of differential projectors that
enforce the gauge condition on each coordinate $x^{\mu}$ and
$z^{\mu}$, acting on a scalar structure function. For the spin zero
part this form is,
\begin{equation}
i\Bigl[\mbox{}_{\mu\nu} \Delta^0_{\rho\sigma}\Bigr](x;z) =
\mathcal{P}_{\mu\nu}(x) \times \mathcal{P}_{\rho\sigma}(z)
\Bigl[\mathcal{S}_0(x;z) \Bigr] \; . \label{Spin0}
\end{equation}
The projector $\mathcal{P}_{\mu\nu}$ is,
\begin{equation}
\mathcal{P}_{\mu\nu} \equiv D_{\mu} D_{\nu} + \frac{g_{\mu\nu}}{D
\!-\!2} \Bigl[ \square \!+\! 2 (D \!-\! 1) H^2 \Bigr] \; .
\label{spin0op}
\end{equation}
The spin two part is more complicated,
\begin{equation}
i\Bigl[\mbox{}_{\mu\nu} \Delta^2_{\rho\sigma}\Bigr](x;z) = \frac1{4
H^4} \mathbf{P}_{\mu\nu}^{~~\alpha\beta}(x) \times
\mathbf{P}_{\rho\sigma}^{~~\kappa\lambda}(z) \Bigl[
\mathcal{R}_{\alpha\kappa}(x;z) \mathcal{R}_{\beta\lambda}(x;z)
\mathcal{S}_2(x;z)\Bigr] \; . \label{Spin2}
\end{equation}
The projector $\mathbf{P}_{\mu\nu}^{~~\alpha\beta}$ is,
\begin{eqnarray}
\lefteqn{\mathbf{P}_{\mu\nu}^{~~\alpha\beta} = \frac12
\Bigl(\frac{D\!-\!3}{D\!-\!2}\Bigr) \Biggl\{ -\delta^{\alpha}_{(\mu}
\delta^{\beta}_{\nu)} \Bigl[\square \!-\! D H^2\Bigr] \Bigl[ \square
\!-\! 2 H^2\Bigr] + 2 D_{(\mu} \Bigl[ \square \!+\! H^2\Bigr]
\delta^{(\alpha}_{\nu)} D^{\beta)} } \nonumber \\
& & \hspace{.3cm} - \Bigl( \frac{D \!-\!2}{D \!-\!1} \Bigr) D_{(\mu}
D_{\nu)} D^{(\alpha} D^{\beta)} + g_{\mu\nu} g^{\alpha\beta} \Bigl[
\frac{\square^2}{D \!-\! 1} \!-\! H^2 \square \!+\! 2 H^4\Bigr]
\qquad \nonumber \\
& & \hspace{.3cm} -\frac{D_{(\mu} D_{\nu)} }{D \!-\! 1} \Bigl[
\square \!+\! 2 (D \!-\! 1) H^2\Bigr] g^{\alpha\beta}
-\frac{g_{\mu\nu} }{D \!-\! 1} \Bigl[ \square \!+\! 2 (D \!-\! 1)
H^2\Bigr] D^{(\alpha} D^{\beta)} \Biggr\} . \qquad \label{spin2op}
\end{eqnarray}
And $\mathcal{R}_{\alpha\kappa}$ is the mixed derivative of the
length function, normalized to give $\eta_{\alpha\kappa}$ in the
flat space limit,
\begin{equation}
\mathcal{R}_{\alpha\kappa}(x;z) \equiv -\frac1{2 H^2}
\frac{\partial^2 y(x;z)}{\partial x^{\alpha} \partial z^{\kappa}} \;
.
\end{equation}

\section{Expansions for the Structure Functions}

The purpose of this section is to facilitate dimensional
regularization computations by giving explicit expansions for the
two scalar structure functions that appear in expressions
(\ref{Spin0}) and (\ref{Spin2}), $\mathcal{S}_0(x;z)$ and
$\mathcal{S}_2(x;z)$. Because the differential operators
(\ref{spin0op}) and (\ref{spin2op}) can be simplified when one knows
something about the function upon which they act, it is convenient
to decompose each structure function into a de Sitter invariant part
which depends only upon $y(x;z)$ and a de Sitter breaking part that
also depends upon the scale factors $a_x$ and $a_z$,
\begin{equation}
\mathcal{S}_i(x;z) = S_i(y) + \delta S_i(a_x,a_z,y) \; .
\end{equation}
The fundamental expressions for each structure function involve the
scalar propagators $i\Delta_b(x;z)$ described in the previous
section. Four choices of the index $b$ occur so frequently that they
have merited a special notation,
\begin{equation}
b_B \equiv \Bigl( \frac{D \!-\! 3}2\Bigr) \;\; , \;\; b_A \equiv
\Bigl( \frac{D \!-\! 1}2\Bigr) \;\; , \;\; b_W \equiv \Bigl( \frac{D
\!+\! 1}2\Bigr) \;\; , \;\; b_M \equiv \sqrt{\frac{(D\!-\!1) (D
\!+\! 7)}4} \; . \label{bdefs}
\end{equation}
The $B$-type propagator corresponds to a positive mass-squared of
$M_S^2 = (D - 2) H^2$ and is de Sitter invariant; the others all
have de Sitter breaking parts. The naive mode sums for the $A$-type
propagator (with $M_S^2 = 0$) and the $W$-type propagator (with
$M_S^2 = -D H^2$) both harbor logarithmic infrared divergences,
which require combining the de Sitter breaking corrections
$\Delta^{\rm IR}_{\nu}(x;z)$ with divergent Gamma functions in
$\Delta^{\rm dS}_{\nu}(x;z)$ before taking the limit $\nu = b_A$ or
$\nu = b_W$. The infrared divergences in the $M$-type propagator
(with $M_S^2 = -2(D-1) H^2$) are of the power law type which do not
require such care.

\subsection{Spin Zero Part}

The spin zero structure function can be successively decomposed into
more and more explicit combinations of scalar propagators,
\begin{eqnarray}
\mathcal{S}_0(x;z) & = & -2 \Bigl( \frac{D \!-\! 2}{D \!-\! 1}\Bigr)
\, i\Delta_{WMM}(x;z) \; , \\
& = & \frac{2}{(D \!-\! 1) H^2} \Bigl[ i\Delta_{MM}(x;z) \!-\! i
\Delta_{MW}(x;z) \Bigr] \; , \\
& = & \frac1{(D \!-\!1)H^4} \Biggl\{ -\frac1{b_M} \Bigl(\frac{
\partial i\Delta_{b}}{\partial b}\Bigr)_{\!\!M} + \frac{2 i
\Delta_{M}}{D\!-\!2} - \frac{2 i \Delta_W} {D\!-\!2} \Biggr\} .
\qquad
\end{eqnarray}
Owing to the factors of $\square + 2 (D-1) H^2$ in the spin zero
projectors (\ref{spin0op}) it is also desirable to give explicit
results for $i\Delta_{MW}(x;z)$ and $i\Delta_{W}(x;z)$,
\begin{eqnarray}
\lefteqn{i\Bigl[\mbox{}_{\mu\nu} \Delta^0_{\rho\sigma}\Bigr](x;z) =
\frac{D^4 \mathcal{S}_0(x;z)}{Dx^{\mu} Dx^{\nu} Dz^{\rho}
Dz^{\sigma}} - \frac{2 g_{\mu\nu}(x)}{D \!-\! 1} \, \frac{D^2
i\Delta_{MW}}{Dz^{\rho} Dz^{\sigma}} } \nonumber \\
& & \hspace{4cm} - \frac{2 g_{\rho\sigma}(z)}{D \!-\! 1} \,
\frac{D^2 i\Delta_{MW}}{Dx^{\mu} Dx^{\nu}} - \frac{2 g_{\mu\nu}(x)
g_{\rho\sigma}(z) i\Delta_{W}}{(D \!-\! 2) (D \!-\! 1)} \; . \qquad
\label{zerodecomp}
\end{eqnarray}
As usual, we decompose them into de Sitter invariant and breaking
parts,
\begin{eqnarray}
i\Delta_{MW}(x;z) & = & MW(y) + \delta MW(a_x,a_z,y) \; , \\
i\Delta_{W}(x;z) & = & W(y) + \delta W(a_x,a_z,y) \; .
\end{eqnarray}

We begin with the de Sitter invariant parts. The function $W(y)$
enters the propagator without any derivatives so its leading
singularity is $(4/y)^{\frac{D}2 -1}$,
\begin{eqnarray}
\lefteqn{W(y) = \frac{H^{D-2}}{(4\pi)^{\frac{D}2}} \Biggl\{
\Gamma\Bigl(\frac{D}2 \!-\!1\Bigr)
\Bigl(\frac{4}{y}\Bigr)^{\frac{D}2 -1} \!+\! \frac{\Gamma(\frac{D}2
\!+\! 2)}{(\frac{D}2 \!-\! 2) (\frac{D}2 \!-\!1)}
\Bigl(\frac{4}{y} \Bigr)^{\frac{D}2-2} } \nonumber \\
& & \hspace{3cm} \!+\! \frac{\Gamma(\frac{D}2 \!+\! 3)}{2 (\frac{D}2
\!-\! 3) (\frac{D}2\!-\!2)} \Bigl(\frac{4}{y} \Bigr)^{\frac{D}2-3}
\!+\! W_1 \!+\!
W_2 \Bigl(\frac{y \!-\!2}4\Bigr) \nonumber \\
& & \hspace{1.5cm} + \sum_{n=2}^{\infty} \Biggl[
\frac{\Gamma(n\!+\!\frac{D}2\!+\!2)
(\frac{y}4)^{n-\frac{D}2+2}}{(n\!-\! \frac{D}2\!+\!2) (n \!-\!
\frac{D}2 \!+\!1) (n \!+\! 1)!} - \frac{\Gamma(n \!+\! D)
(\frac{y}4)^n }{n (n \!-\!1) \Gamma(n \!+\! \frac{D}2)} \Biggr]
\Biggr\} , \qquad \label{DeltaW}
\end{eqnarray}
where the constants $W_1$ and $W_2$ are,
\begin{eqnarray}
W_1 & = & \frac{\Gamma(D\!+\!1)}{\Gamma(\frac{D}2 \!+\!1)} \Biggl\{
\frac{D \!+\!1}{2 D} \Biggr\} , \\
W_2 & = & \frac{\Gamma(D\!+\!1)}{\Gamma(\frac{D}2 \!+\!1)} \Biggl\{
\psi\Bigl(-\frac{D}2\Bigr) - \psi\Bigl(\frac{D\!+\!1}2\Bigr) -
\psi(D \!+\!1) - \psi(1) \Biggr\} .
\end{eqnarray}
The function $MW(y)$ enters the graviton propagator with two
derivatives so its leading singularity is $(4/y)^{\frac{D}2-2}$,
\begin{eqnarray}
\lefteqn{MW(y) = \frac{H^{D-4}}{(4\pi)^{\frac{D}2}} \Biggl\{ -
\frac{\Gamma(\frac{D}2 \!-\! 1)}{\frac{D}2 \!-\! 2}
\Bigl(\frac{4}{y}\Bigr)^{\frac{D}2 - 2} + {\rm constant} } \nonumber
\\
& & \hspace{4cm} + \sum_{n=1}^{\infty} \Biggl[ (MW)^a_n
\Bigl(\frac{y}{4}\Bigr)^{n-\frac{D}2+2} - (MW)^b_n
\Bigl(\frac{y}{4}\Bigr)^{n} \Biggr] \Biggr\} . \qquad
\label{DeltaMW}
\end{eqnarray}
The coefficients $(MW)^a_n$ and $(MW)^b_n$ are,
\begin{eqnarray}
(MW)^a_n & = & \frac{\Gamma(n \!+\! \frac{D}2 \!+\! 2)}{(D \!-\! 2)
(n \!-\! \frac{D}2 \!+\! 2) (n \!-\! \frac{D}2 \!+\! 1) (n\!+\!1)! }
\nonumber \\
& & \hspace{1.5cm} + \frac{\Gamma(\frac{D}2\!-\! 1) \Gamma(1 \!-\!
\frac{D}2)}{2 \Gamma(3 \!-\! \frac{D}2 \!+\! n)}
\frac{\Gamma(\frac32 \!+\! b_M \!+\! n) \Gamma(\frac32 \!-\! b_M
\!+\! n)}{\Gamma(\frac12 \!+\! b_M) \Gamma(\frac12 \!-\! b_M)
(n \!+\! 1)! } \; , \qquad \\
(MW)^b_n & = & \frac{\Gamma(n \!+\! D)}{(D \!-\! 2) \Gamma(n \!+\!
\frac{D}2) n (n \!-\! 1)} \nonumber \\
& & \hspace{1.5cm} + \frac{\Gamma(\frac{D}2\!-\! 1) \Gamma(1 \!-\!
\frac{D}2)}{2 \Gamma(\frac{D}2 \!+\! n)} \frac{\Gamma(b_A \!+\! b_M
\!+\! n) \Gamma(b_A \!-\! b_M \!+\! n)}{\Gamma(\frac12 \!+\! b_M)
\Gamma(\frac12 \!-\! b_M) n! } \; , \qquad
\end{eqnarray}
and we should point out the special definition for $(MW)^b_1$,
\begin{equation}
(MW)^b_1 \equiv \frac{W_2}{D \!-\! 2} + \frac{2 \Gamma(1 \!-\!
\frac{D}2)}{(D \!-\! 2) D} \frac{\Gamma(b_W \!+\! b_M) \Gamma(b_W
\!-\! b_M)}{\Gamma(\frac12 \!+\! b_M) \Gamma(\frac12 \!-\! b_M)} \;
.
\end{equation}
That brings us to $S_0(y)$ which enters the graviton propagator with
four derivatives and accordingly begins with $(4/y)^{\frac{D}2-3}$,
\begin{eqnarray}
\lefteqn{S_0(y) = \Bigl( \frac{D\!-\! 2}{D \!-\! 1} \Bigr)
\frac{H^{D-6}}{(4\pi)^{\frac{D}2}} \Biggl\{ - \frac{\Gamma(\frac{D}2
\!-\! 1)}{(\frac{D}2 \!-\! 3) (\frac{D}2 \!-\! 2)} \Bigl(
\frac{4}{y} \Bigr)^{\frac{D}2 - 3} + {\rm constant} - (S_0)^b_1
\Bigl(\frac{y}4\Bigr) } \nonumber \\
& & \hspace{5cm} + \sum_{n=2}^{\infty} \Biggl[ (S_0)^a_n
\Bigl(\frac{y}{4}\Bigr)^{n-\frac{D}2+2} - (S_0)^b_n
\Bigl(\frac{y}{4}\Bigr)^{n} \Biggr] \Biggr\} . \qquad
\label{DeltaS0}
\end{eqnarray}
The coefficients are,
\begin{eqnarray}
(S_0)^a_n & = & -\frac{2 (WM)^a_n}{D \!-\! 2} +
\frac{\Gamma(\frac{D}2\!-\! 1) \Gamma(1 \!-\! \frac{D}2)}{2 \Gamma(3
\!-\! \frac{D}2 \!+\! n)} \frac{\Gamma(\frac32 \!+\! b_M \!+\! n)
\Gamma(\frac32 \!-\! b_M \!+\! n)}{b_M \Gamma(\frac12 \!+\!
b_M) \Gamma(\frac12 \!-\! b_M) (n \!+\! 1)! } \nonumber \\
& & \hspace{-1cm} \times \Biggl\{ \psi\Bigl( \frac32 \!+\! b_M \!+\!
n\Bigr) - \psi\Bigl( \frac32 \!-\! b_M \!+\! n\Bigr) - \psi\Bigl(
\frac12 \!+\! b_M\Bigr) + \psi\Bigl( \frac12 \!-\! b_M\Bigr)
\Biggr\} , \qquad \\
(S_0)^b_n & = & -\frac{2 (WM)^b_n}{D \!-\! 2} + \frac{\Gamma(
\frac{D}2 \!-\! 1) \Gamma(1 \!-\! \frac{D}2)}{2 \Gamma(\frac{D}2
\!+\! n)} \frac{\Gamma(b_A \!+\! b_M \!+\! n) \Gamma(b_A \!-\! b_M
\!+\! n)}{b_M \Gamma(\frac12 \!+\! b_M)
\Gamma(\frac12 \!-\! b_M) n! } \nonumber \\
& & \hspace{-1cm} \times \Biggl\{ \psi\Bigl(b_A \!+\! b_M \!+\!
n\Bigr) - \psi\Bigl(b_A \!-\! b_M \!+\! n\Bigr) - \psi\Bigl( \frac12
\!+\! b_M\Bigr) + \psi\Bigl( \frac12 \!-\! b_M\Bigr) \Biggr\} .
\qquad
\end{eqnarray}

An important point to note about the $M$-type contributions is that
the infinite series terms do not vanish in $D=4$ dimensions, in
spite of the fact that the coefficients of the
$(\frac{y}4)^{n-\frac{D}2+2}$ and the $(\frac{y}{4})^n$ terms agree
for $D=4$. This is because both are multiplied by the Gamma function
$\Gamma(1-\frac{D}2)$ which diverges for $D=4$.

Each of the de Sitter breaking parts is the sum of three terms:
\begin{enumerate}
\item{A power of $a_x a_z$;}
\item{A power of $a_x a_z$ times $(y-2$); and}
\item{A power of $a_x a_z$ times $(\frac{a_x}{a_z} +
\frac{a_z}{a_x})$.}
\end{enumerate}
The two primitive contributions are \cite{MTW1},
\begin{eqnarray}
\delta W & = & k \Biggl\{ 4 b_A^2 a_x a_z - b_A \ln(a_x a_z) (y
\!-\! 2) - \Bigl(\frac{a_x}{a_z} \!+\! \frac{a_z}{a_x}\Bigr)
\Biggr\} \; , \qquad \\
\delta M & = & k_M \Biggl\{ \frac{ (a_x a_z)^{b_M - b_A}}{b_M \!-\!
b_A} - \frac{ (a_x a_z)^{b_M - b_A - 1}}{b_M \!-\! b_A \!-\! 1}
\times \frac{(y \!-\! 2)}{4 b_A} \nonumber \\
& & \hspace{5cm} - \frac{ (a_x a_z)^{b_M - b_A - 1}}{4 b_A (b_M
\!-\! 1)} \times \Bigl( \frac{a_x}{a_z} \!+\! \frac{a_z}{a_x}\Bigr)
\Biggr\} \; . \qquad
\end{eqnarray}
Recall $b_A$ and $b_M$ from (\ref{bdefs}). The constants $k$ and
$k_M$ are,
\begin{equation}
k \equiv \frac{H^{D-2}}{(4\pi)^{\frac{D}2}} \, \frac{\Gamma(D \!-\!
1)}{ \Gamma(\frac{D}2)} \qquad , \qquad k_M \equiv
\frac{H^{D-2}}{(4\pi)^{\frac{D}2}} \frac{ \Gamma(b_M)
\Gamma(2b_M)}{\Gamma(b_A) \Gamma(b_M \!+\! \frac12)} \; .
\label{kdefs}
\end{equation}

It is useful to represent spacetime dependence using $y(x;z)$ and
two de Sitter breaking combinations of the scale factors,
\begin{equation}
u \equiv \ln(a_x a_z) \qquad , \qquad v \equiv
\ln\Bigl(\frac{a_x}{a_z}\Bigr) \; .
\end{equation}
Each of the three de Sitter breaking contributions takes the form,
\begin{equation}
f(u,v,y) = f_1(u) + f_2(u) \times (y \!-\! 2) + f_3(u) \times
\cosh(v) \; .
\end{equation}
Hence we need only give the functions $f_i(u)$ for each of the three
combinations which enter expression (\ref{zerodecomp}). The simplest
is,
\begin{equation}
\frac{-2}{(D \!-\! 2) (D \!-\! 1 )} \, \frac{\delta W}{H^4} =
f_1^{W}(u) + f_2^{W}(u) \times (y \!-\! 2) + f_3^{W}(u) \times
\cosh(v) \; ,
\end{equation}
whose the coefficient functions are,
\begin{equation}
f_1^{W} = \frac{k}{H^4} \times -2 \Bigl(\frac{D \!-\! 1}{D \! -\!
2}\Bigr) e^{u} \;\; , \;\; f_2^{W} = \frac{k}{H^4} \times \frac{u}{D
\!-\! 2} \;\; , \; \; f_3 = \frac{k}{H^4} \times \frac{4}{(D \!-\!
2)( D \!-\! 1)} \; . \label{fW}
\end{equation}
The next simplest combination is,
\begin{equation}
\frac{-2}{(D \!-\! 1)} \frac{\delta MW}{H^2} = f_1^{MW}(u) +
f_2^{MW}(u) \times (y \!-\! 2) + f_3^{MW}(u) \times \cosh(v) \; .
\end{equation}
Its coefficient functions involve (\ref{fW}),
\begin{eqnarray}
f_1^{MW}(u) & = & f_1^{W}(u) + \frac{k_M}{H^4} \times \frac{e^{(b_M
- b_A) u}}{(D \!-\! 2) b_A (b_M \!-\! b_A)} \; , \label{fMW1} \\
f_2^{MW}(u) & = & f_2^{W}(u) + \frac{k_M}{H^4} \times \frac{-e^{(b_M
- b_W) u}}{4 (D \!-\! 2) b_A^2 (b_M \!-\! b_W)} \; , \label{fMW2} \\
f_3^{MW}(u) & = & f_3^{W}(u) + \frac{k_M}{H^4} \times \frac{-e^{(b_M
- b_W) u}}{2 (D \!-\! 2) b_A^2 (b_M \!-\! 1)} \; . \label{fMW3}
\end{eqnarray}
Of course the most complicated is $\delta S_0$ itself,
\begin{equation}
\delta S_0 = f_1^{S_0}(u) + f_2^{S_0}(u) \times (y \!-\! 2) +
f_3^{S_0}(u) \times \cosh(v) \; . \label{deltaS0}
\end{equation}
Its coefficient functions involve the constant $C_M \equiv
2\psi(b_M) + 2 \ln(2)$,
\begin{eqnarray}
f_1^{S_0}(u) & = & f_1^{MW}(u) + \frac{k_M}{H^4} \times
\frac{e^{(b_M - b_A) u}}{2 b_A b_M} \times \Biggl\{ \frac{-C_M \!-\!
u}{b_M \!-\! b_A} \!+\! \frac1{(b_M \!-\! b_A)^2}
\Biggr\} , \label{fS01} \\
f_2^{S_0}(u) & = & f_2^{MW}(u) + \frac{k_M}{H^4} \times
\frac{e^{(b_M - b_W) u}}{8 b_A^2 b_M} \times \Biggl\{ \frac{C_M
\!+\! u}{b_M \!-\! b_W} \!-\! \frac1{(b_M \!-\! b_W)^2} \Biggr\} ,
\qquad \label{fS02} \\
f_3^{S_0}(u) & = & f_3^{MW}(u) + \frac{k_M}{H^4} \times
\frac{e^{(b_M - b_W) u}}{4 b_A^2 b_M} \times \Biggl\{ \frac{C_M
\!+\! u}{b_M \!-\! 1} \!-\! \frac1{(b_M \!-\! 1)^2} \Biggr\} .
\label{fS03}
\end{eqnarray}

\subsection{Spin Two Part}

The spin two structure function can also be decomposed into single
index propagators and their derivatives,
\begin{eqnarray}
\lefteqn{\mathcal{S}_2(x;z) = \frac{32}{(D \!-\! 3)^2} \Biggl\{
i\Delta_{AAA}(x;z) \!-\! 2 i\Delta_{AAB}(x;z) \!+\!
i\Delta_{ABB}(x;z)\Biggr\} , } \\
& & \hspace{0cm} = \frac{32}{(D \!-\! 3)^2 H^4} \Biggl\{ \frac1{2
(D\!-\!1)^2} \Bigl(\frac{\partial^2 i\Delta_b}{\partial b^2}
\Bigr)_{\!\!A} - \frac{(2D \!-\! 3) D}{(D \!-\! 2) (D \!-\! 1)^3}
\Bigl(\frac{\partial i\Delta_b}{\partial b}\Bigr)_{\!\!A} \nonumber \\
& & \hspace{3cm} + \frac{3 i\Delta_A}{(D \!-\! 2)^2} - \frac1{(D
\!-\! 3) (D\!-\!2)} \Bigl(\frac{\partial i\Delta_b}{\partial
b}\Bigr)_{\!\!B} - \frac{3 i\Delta_B}{(D \!-\! 2)^2} \Biggr\} .
\qquad
\end{eqnarray}

The spin two structure function can be acted on by up to eight
derivatives in the graviton propagator so its leading singularity is
$(\frac{4}{y})^{\frac{D}2-5}$,
\begin{eqnarray}
\lefteqn{ S_2(y) = 32 \Bigl( \frac{D \!-\! 2}{D \!-\! 3}\Bigr)^2
\frac{H^{D-6}}{(4\pi)^{\frac{D}2}} \Biggl\{ \frac{ \Gamma(\frac{D}2
\!-\! 1) (\frac{4}{y})^{\frac{D}2-5} }{4! (\frac{D}2 \!-\! 5)
(\frac{D}2 \!-\! 4) (\frac{D}2 \!-\! 3) (\frac{D}2 \!-\! 2)}
+ {\rm constant} } \nonumber \\
& & \hspace{4cm} + \sum_{n=4}^{\infty} (S_2)^a_n \times
\Bigl(\frac{y}{4}\Bigr)^{n-\frac{D}2+2} - \sum_{n=1}^{\infty}
(S_2)^b_n \times \Bigl(\frac{y}4\Bigr)^n \Biggr\} . \qquad
\label{DeltaS2}
\end{eqnarray}
To specify the coefficients $(S_2)^a_n$ and $(S_2)^b_n$ it is useful
to make the preliminary definitions,
\begin{eqnarray}
\Psi^a_n(\nu) & \equiv & \psi\Bigl(\frac32 \!+\! \nu \!+\! n\Bigr) -
\psi\Bigl(\frac12 \!+\! \nu\Bigr) - \psi\Bigl(\frac32 \!-\! \nu
\!+\! n\Bigr) + \psi\Bigl(\frac12 \!-\! \nu\Bigr) \; , \qquad \\
\Psi^b_n(\nu) & \equiv & \psi\Bigl(b_A \!+\! \nu \!+\! n\Bigr) -
\psi\Bigl(\frac12 \!+\! \nu\Bigr) - \psi\Bigl(b_A \!-\! \nu \!+\!
n\Bigr) + \psi\Bigl(\frac12 \!-\! \nu\Bigr) \; . \qquad
\end{eqnarray}
The coefficients in (\ref{DeltaS2}) are,
\begin{eqnarray}
\lefteqn{ (S_2)^a_n = \frac{-\Gamma(\frac{D}2 \!+\! 1 \!+\! n)}{(D
\!-\! 2)^2 (2 \!-\! \frac{D}2 \!+\! n) (n \!+\!1)!} \Biggl\{
\frac{\Psi_n^{a\prime}(b_A) \!+\! [\Psi_n^a(b_A)]^2}{ 2 (D \!-\!
1)^2} - \frac{ \Psi^a_n(b_A)}{(D \!-\! 1)^3} } \nonumber \\
& & \hspace{.7cm} - \frac{2 \Psi^a_n(b_A)}{(D \!-\! 2) (D \!-\! 1)}
+ \frac{6}{(D \!-\! 2)^2} - \frac{[3 (D \!-\! 3) \!+\! (\frac{D}2
\!-\! 2 \!-\! n) \Psi^a_n(b_B)]}{(\frac{D}2 \!+\!
n)(D \!-\! 3) (D \!-\! 2)} \Biggr\} \; , \qquad \\
\lefteqn{ (S_2)^b_n = -\frac{\Gamma(D \!-\! 1 \!+\! n)}{(D \!-\!
2)^2 n \Gamma(\frac{D}2 \!+\! n)} \Biggl\{
\frac{\Psi_n^{b\prime}(b_A) \!+\! [\Psi_n^b(b_A)]^2}{ 2 (D \!-\!
1)^2} - \frac{ \Psi^b_n(b_A)}{ (D \!-\! 1)^3} } \nonumber \\
& & \hspace{1.5cm} - \frac{2 \Psi^b_n(b_A)}{(D \!-\! 2) (D \!-\! 1)}
+ \frac{6}{(D \!-\! 2)^2} -\frac{[3(D \!-\! 3) \!-\! n
\Psi^b_n(b_B)]}{(D \!-\! 2 \!+\! n)(D \!-\! 3) (D \!-\! 2)} \Biggr\}
\; . \qquad \label{S2bn}
\end{eqnarray}
It might be worth noting that $\Psi^a_n(\nu)$ can be simplified,
\begin{equation}
\Psi^a_n(\nu) = -\sum_{m=0}^n \frac{2 \nu}{(m \!+\! \nu \!+\!
\frac12) (m \!-\! \nu \!+\! \frac12)} \; .
\end{equation}
This permits a more explicit form of $(S_2)^a_n$,
\begin{eqnarray}
\lefteqn{ (S_2)^a_n = \frac{-\Gamma(\frac{D}2 \!+\! 1 \!+\! n)}{(D
\!-\! 2)^2(2 \!-\! \frac{D}2 \!+\! n) (n \!+\! 1)!} \Biggl\{
\sum_{m=0}^{n-1} \frac1{(\frac{D}2 \!+\! m) (\frac{D}2 \!-\! 1 \!-\!
m)} } \nonumber \\
& & \hspace{-.5cm} \times \!\! \sum_{\ell = m+1}^n \frac1{(\frac{D}2
\!+\! \ell) (\frac{D}2 \!-\! 1 \!-\! \ell)} - \frac2{D\!-\!2}
\sum_{m=0}^n \frac1{(\frac{D}2 \!+\! m) (\frac{D}2 \!-\! 1 \!-\! m)}
+ \frac6{(D \!-\! 2)^2}
\nonumber \\
& & \hspace{.8cm} - \frac3{(D \!-\! 2) (\frac{D}2 \!+\! n)} -
\frac{(\frac{D}2 \!-\! 2 \!-\! n)}{(D \!-\! 2) (\frac{D}2 \!+\! n)}
\sum_{m=0}^n \frac1{(\frac{D}2 \!+\! m \!-\! 1) (\frac{D}2 \!-\! 2
\!-\! m)} \Biggr\} . \qquad
\end{eqnarray}

In comparison with its spin zero cousin the de Sitter breaking of
the spin two structure function is simple. It derives entirely from
the $A$ type propagator and its derivatives. Further, its spacetime
dependence is limited to powers of $\ln(4 a_x a_z)$, with no
dependence upon either $y$ or $(\frac{a_x}{a_z} + \frac{a_z}{a_x})$.
Our result for it is,
\begin{eqnarray}
\lefteqn{ \delta S_2 = \frac{32 k}{(D \!-\! 3)^2 H^4} \Biggl\{
\frac{\ln^3(4 a_x a_z)}{6 (D \!-\! 1)^2} + \Biggl[ \frac{\psi(b_A)}{
(D \!-\! 1)^2} -\frac{(D \!-\! \frac32) D}{(D \!-\! 2) (D
\!-\! 1)^3} \Biggr] \ln^2(4 a_x a_z) } \nonumber \\
& & \hspace{.5cm} + \Biggl[ \frac{\psi'(b_A) \!+\! 2 \psi^2(b_A)}{(D
\!-\! 1)^2} - \frac{2 (2D \!-\! 3) D \psi(b_A)}{(D \!-\! 2) (D
\!-\!1)^3} + \frac3{(D \!-\! 2)^2} \Biggr] \ln(4 a_x a_z) \Biggr\} .
\qquad \label{deltaS2}
\end{eqnarray}

\section{de Sitter Invariant Tensor Structure}

\begin{table}

\vbox{\tabskip=0pt \offinterlineskip
\def\tablerule{\noalign{\hrule}}
\halign to390pt {\strut#& \vrule#\tabskip=1em plus2em& \hfil#&
\vrule#& \hfil#\hfil&\vrule#\tabskip=0pt\cr \tablerule
\omit&height4pt&\omit&&\omit&\cr &&\omit\hidewidth k
&&\omit\hidewidth $[\mbox{}_{\mu\nu}
\mathcal{T}^k_{\rho\sigma}](x;z)$ \hidewidth&\cr
\omit&height4pt&\omit&&\omit&\cr \tablerule
\omit&height2pt&\omit&&\omit&\cr && 1 && $\frac{\partial^2
y}{\partial x^{\mu} \partial z^{(\rho}} \frac{\partial^2 y}{\partial
z^{\sigma)} \partial x^{\nu}}$ &\cr \omit&height2pt&\omit&&\omit&\cr
\tablerule \omit&height2pt&\omit&&\omit&\cr && 2 && $\frac{\partial
y}{\partial x^{(\mu}} \frac{\partial^2 y}{\partial x^{\nu)} \partial
z^{(\rho}} \frac{\partial y}{\partial z^{\sigma)}}$ &\cr
\omit&height2pt&\omit&&\omit&\cr \tablerule
\omit&height2pt&\omit&&\omit&\cr && 3 && $\frac{\partial y}{\partial
x^{\mu}} \frac{\partial y}{\partial x^{\nu}} \frac{\partial
y}{\partial z^{\rho}} \frac{\partial y}{\partial z^{\sigma}}$ &\cr
\omit&height2pt&\omit&&\omit&\cr \tablerule
\omit&height2pt&\omit&&\omit&\cr && 4 && $H^2 [g_{\mu\nu}(x)
\frac{\partial y}{\partial z^{\rho}} \frac{\partial y}{\partial
z^{\sigma}} + \frac{\partial y}{\partial x^{\mu}} \frac{\partial
y}{\partial x^{\nu}} g_{\rho\sigma}(z)]$ &\cr
\omit&height2pt&\omit&&\omit&\cr \tablerule
\omit&height2pt&\omit&&\omit&\cr && 5 && $H^4 g_{\mu\nu}(x)
g_{\rho\sigma}(z)$ &\cr \omit&height2pt&\omit&&\omit&\cr
\tablerule}}

\caption{Tensor basis for representing the de Sitter invariant part
of $i[\mbox{}_{\mu\nu}\Delta_{\rho\sigma}](x;z)$. See
Table~\ref{nondStens} for the nine de Sitter breaking basis
tensors.}

\label{dStens}

\end{table}

In this section we express the de Sitter invariant part of the
propagator in terms of the five invariant basis tensors employed
previously to represent the graviton self-energy \cite{PW}, which
are given in Table ~\ref{dStens}. That is, we represent the de
Sitter invariant contributions to the spin zero and the spin two
parts of the graviton propagator as linear combinations of the five
basis tensors,
\begin{equation}
i \Bigl[\mbox{}_{\mu\nu} \Delta^{\rm dS}_{\rho\sigma} \Bigr](x;z) =
\sum_{k=1}^5 \Bigl[\mathcal{C}^k_0(y) + \mathcal{C}^k_2(y) \Bigr]
\times \Bigl[\mbox{}_{\mu\nu} \mathcal{T}^k_{\rho\sigma}\Bigr](x;z)
\; .
\end{equation}
Each of the combination coefficients $\mathcal{C}^k_{0,2}(y)$ can be
expressed in terms of the de Sitter invariant parts given in the
previous section, and their derivatives. There does not seem to be
any point to giving explicit series expansions for each coefficient.

For the spin zero case a useful preliminary result is,
\begin{equation}
\frac{D^2 F(y)}{Dx^{\mu} Dx^{\nu}} = H^2 g_{\mu\nu}(x) (2 \!-\! y)
F'(y) + \frac{\partial y}{\partial x^{\mu}} \, \frac{\partial
y}{\partial x^{\nu}} \, F''(y) \; .
\end{equation}
Now use this in (\ref{zerodecomp}) to find the five spin zero
coefficients,
\begin{eqnarray}
\mathcal{C}_0^1(y) & = & 2 S_0''(y) \; , \label{C01} \\
\mathcal{C}_0^2(y) & = & 4 S_0'''(y) \; , \label{C02} \\
\mathcal{C}_0^3(y) & = & S_0''''(y) \; , \label{C03} \\
\mathcal{C}_0^4(y) & = & - 2 S_0''(y) + (2 \!-\! y) S_0'''(y)
-\frac2{D \!-\! 1} \, \frac{MW''(y)}{H^2} \; , \qquad \label{C04} \\
\mathcal{C}_0^5(y) & = & -(2 \!-\! y) S_0'(y) + (2 \!-\! y)^2
S_0''(y) \nonumber \\
& & \hspace{2cm} -\frac4{D \!-\! 1} \, \frac{(2 \!-\! y)
MW'(y)}{H^2} - \frac2{(D \!-\! 2) (D \!-\! 1)} \, \frac{W(y)}{H^4}
\; . \qquad \label{C05}
\end{eqnarray}

The de Sitter invariant contribution from the spin two part is,
\begin{equation}
i\Bigl[\mbox{}_{\mu\nu} \Delta^{\rm dS,2}_{\rho\sigma}\Bigr](x;z) =
\frac1{4 H^4} \mathbf{P}_{\mu\nu}^{~~\alpha\beta}(x) \times
\mathbf{P}_{\rho\sigma}^{~~\kappa\lambda}(z) \Bigl[
\mathcal{R}_{\alpha\kappa}(x;z) \mathcal{R}_{\beta\lambda}(x;z)
S_2(y)\Bigr] \; .
\end{equation}
From expression (\ref{spin2op}) we see that the transverse-traceless
projector $\mathbf{P}_{\mu\nu}^{~~\alpha\beta}(x)$ contains four
derivative operators. We must therefore work out what happens when
a derivative acts on two factors of $\mathcal{R}$, for example,
\begin{equation}
\frac{D}{D x^{\alpha}} \Bigl[ \Bigl(
\mathcal{R}^{\alpha}_{~\kappa} \mathcal{R}_{\nu \lambda} \!+\!
\mathcal{R}_{\nu \kappa} \mathcal{R}^{\alpha}_{~\lambda} \Bigr)
F\Bigr] = (D \!+\! 1) \mathcal{R}_{\nu (\kappa} \frac{\partial
y}{\partial z^{\lambda)}} F + 2 \mathcal{R}_{\nu (\kappa}
\mathcal{R}^{\alpha}_{~\lambda)} \frac{D F}{Dx^{\alpha}} \; .
\end{equation}
It turns our that 11 similar identities are needed to act the first
projector; they are given in the Appendix. With these identities it
is straightforward but tedious to show,
\begin{eqnarray}
\lefteqn{ \hspace{-.3cm} 2 \Bigl(\frac{D \!-\! 2}{D \!-\! 3}\Bigr)
\mathbf{P}_{\mu\nu}^{~~\alpha\beta}(x) \Bigl[
\mathcal{R}_{\alpha\kappa} \mathcal{R}_{\beta\lambda} F\Bigr] =
(D\!-\!2)(D\!+\!1) \Biggl\{ H^2 \mathcal{R}_{\mu (\kappa}
\frac{\partial y}{\partial z^{\lambda)}} \frac{DF}{Dx^{\nu}} }
\nonumber \\
& & \hspace{-.5cm} - \frac1{D \!-\! 1} \mathcal{R}^{\alpha}_{~
(\kappa} \frac{\partial y}{\partial z^{\lambda)}} \frac{D^3
F}{Dx^{\mu} Dx^{\nu} Dx^{\alpha}} + \frac1{D\!-\! 1}
\mathcal{R}_{\mu (\kappa} \frac{\partial y}{\partial z^{\lambda)}}
\frac{D}{Dx^{\nu}} \square_x F \Biggr\} \nonumber \\
& & \hspace{-.5cm} + \frac{(D\!-\!2)(D \!+\! 1)}{D \!-\! 1} \Biggl\{
H^2 g_{\kappa\lambda}(z) \frac{D^2F}{Dx^{\mu} Dx^{\nu}} -
\Bigl(\frac{D\!+\!1}4\Bigr) \frac{\partial y}{\partial z^{\kappa}}
\frac{\partial y}{\partial z^{\lambda}} \frac{D^2F}{Dx^{\mu}
Dx^{\nu}} \nonumber \\
& & \hspace{-.5cm} - H^2 \mathcal{R}_{\mu \kappa} \mathcal{R}_{\nu
\lambda} \square_x F + \frac14 g_{\mu\nu}(x) \frac{\partial
y}{\partial z^{\kappa}} \frac{\partial y}{\partial z^{\lambda}}
\square_x F \Biggr\} + 2 \Bigl(\frac{D\!-\!2}{D\!-\!1}\Bigr) \Biggl\{
D H^2 \mathcal{R}_{\mu (\kappa} \mathcal{R}^{\alpha}_{~ \lambda)}
\nonumber \\
& & \hspace{-.5cm} \times \frac{D^2F}{Dx^{\nu} Dx^{\alpha}} - H^2
g_{\mu\nu}(x) \mathcal{R}^{\alpha}_{~(\kappa} \mathcal{R}^{
\beta}_{~\lambda)} \frac{D^2F}{Dx^{\alpha} Dx^{\beta}} \Biggr\}
-\Biggl\{ \Bigl(\frac{D\!-\!2}{D \!-\! 1}\Bigr)
\mathcal{R}^{\alpha}_{~(\kappa} \mathcal{R}^{\beta}_{~\lambda)}
\nonumber \\
& & \hspace{-.5cm} \times \frac{D^4F}{Dx^{\alpha} Dx^{\beta}
Dx^{\mu} Dx^{\nu}} - 2 \mathcal{R}_{\mu (\kappa}
\mathcal{R}^{\alpha}_{~\lambda)} \frac{D^2}{Dx^{\nu} Dx^{\alpha}}
\square_x F +\frac1{D \!-\!1} g_{\mu\nu}(x) \mathcal{R}^{
\alpha}_{~(\kappa} \mathcal{R}^{\beta}_{~\lambda)} \nonumber
\\
& & \hspace{-.5cm} \times \frac{D^2}{D x^{\alpha} Dx^{\beta}}
\square_x F \Biggr\} - \Biggl\{ \mathcal{R}_{\mu \kappa}
\mathcal{R}_{\nu \lambda} \square_x^2 F -
\frac{g_{\kappa\lambda}(z)}{D\!-\! 1} \Bigl[ g_{\mu\nu} \square_x -
\frac{D^2}{Dx^{\mu} Dx^{\nu}} \Bigr] \square_x F \nonumber \\
& & \hspace{2.5cm} +\frac1{4 (D\!-\!1) H^2} \frac{\partial
y}{\partial z^{\kappa}} \frac{\partial y}{\partial z^{\lambda}}
\Bigl[ g_{\mu\nu}(x) \square_x - \frac{D^2}{Dx^{\mu} Dx^{\nu}}
\Bigr] \square_x F \Biggr\} . \label{1stP}
\end{eqnarray}

Our result (\ref{1stP}) can be greatly simplified by taking account
of two facts:
\begin{itemize}
\item{We must still act the operator $\mathbf{P}_{\rho\sigma}^{
~~\kappa\lambda}(z)$, which annihilates any longitudinal or trace
term; and}
\item{The function $F$ depends only upon $y$.}
\end{itemize}
The first fact means that we can neglect total derivatives of
$D/D z^{\kappa}$ or $D/D z^{\lambda}$, and also any term containing
$g_{\kappa\lambda}(z)$. For example, we can write,
\begin{equation}
\frac{\partial y}{\partial z^{\kappa}}
\frac{\partial y}{\partial z^{\lambda}} = 4 H^2 g_{\kappa\lambda}(z)
-4 H^2 \mathcal{R}^{\alpha}_{~\kappa} \mathcal{R}_{\alpha \lambda}
\Longrightarrow -4 H^2 \mathcal{R}^{\alpha}_{~\kappa}
\mathcal{R}_{\alpha \lambda} \; .
\end{equation}
The second fact means that we can trade derivatives of $x$ and $z$,
\begin{equation}
\frac{\partial y}{\partial z^{\lambda}} \frac{D F(y)}{D x^{\nu}} =
\frac{\partial y}{\partial z^{\lambda}}
\frac{\partial y}{\partial x^{\nu}} F'(y) =
\frac{\partial y}{\partial x^{\nu}} \frac{D F(y)}{D z^{\lambda}} \; .
\end{equation}
Exploiting the two facts together allows many simplifications,
for example,
\begin{equation}
\mathcal{R}_{\mu \kappa} \frac{\partial y}{\partial z^{\lambda}}
\frac{D F}{D x^{\nu}} \Longrightarrow 2 H^2 \mathcal{R}_{\mu \kappa}
\mathcal{R}_{\nu \lambda} F \; .
\end{equation}
Eight identities of this type are summarized in the Appendix.
When these relations (\ref{SID1}-\ref{SID8}) are used we get,
\begin{eqnarray}
\lefteqn{ \mathbf{P}_{\mu\nu}^{~~\alpha\beta}(x) \times
\mathbf{P}_{\rho\sigma}^{~~\kappa\lambda}(z) \Bigl[
\mathcal{R}_{\alpha \kappa} \mathcal{R}_{\beta\lambda} F(y) \Bigr] }
\nonumber \\
& & \hspace{2cm} = -\frac12 \Bigl( \frac{D \!-\! 3}{D \!-\! 2}\Bigr)
\mathbf{P}_{\rho\sigma}^{~~\kappa\lambda}(z) \Biggl[
\mathcal{R}_{\mu \kappa} \mathcal{R}_{\nu \lambda} \square \Bigl[
\square \!-\! (D \!-\! 2) H^2\Bigr] F(y) \Biggr] \; . \qquad
\label{1stPsimp}
\end{eqnarray}

It is straightforward to act the final projector by combining
expressions (\ref{1stP}) and (\ref{1stPsimp}). First we define the
function $G(y)$ as,
\begin{eqnarray}
G(y) & \equiv & -\frac14 \frac{\square}{H^2} \Bigl[
\frac{\square}{H^2} \!-\! (D \!-\! 2)\Bigr] S_2(y) \; , \qquad \\
& = & -\frac14 (4y \!-\! y^2)^2 S_2'''' - \Bigl( \frac{D \!+\!
2}2\Bigr) (2 \!-\! y) (4 y \!-\! y^2) S_2''' + \frac{3 D}{4} (4y
\!-\! y^2) S_2'' \nonumber \\
& & \hspace{3cm} - \frac{D (D \!+\! 2)}{4} (2 \!-\! y)^2 S_2'' +
\frac{D (D \!-\! 1)}{2} (2 \!-\! y) S_2' \; . \qquad \label{Gdef}
\end{eqnarray}
Interchanging $x$ and $z$ and their respective index groups then
allows us to read off the result from (\ref{1stP}),
\begin{eqnarray}
\lefteqn{4 \Bigl(\frac{D \!-\! 2}{D \!-\! 3}\Bigr)^2 i \Bigl[
\mbox{}_{\mu\nu} \Delta^{\rm dS,2}_{\rho\sigma} \Bigr](x;z) =
(D\!-\!2)(D\!+\!1) \Biggl\{ H^2 \frac{\partial y}{\partial x^{(\mu}}
\mathcal{R}_{\nu) (\rho}
\frac{DG}{Dz^{\sigma)}} } \nonumber \\
& & \hspace{0cm} - \frac1{D \!-\! 1} \frac{\partial y}{\partial
x^{(\mu}} \mathcal{R}_{\nu)}^{~\alpha} \frac{D^3 G}{Dz^{(\rho}
Dz^{\sigma)} Dz^{\alpha}} + \frac1{D\!-\! 1} \frac{\partial
y}{\partial x^{(\mu}} \mathcal{R}_{\nu) (\rho}
\frac{D}{Dz^{\sigma)}} \square G \Biggr\} \nonumber \\
& & \hspace{0cm} + \frac{(D\!-\!2)(D \!+\! 1)}{D \!-\! 1} \Biggl\{
H^2 g_{\mu\nu}(x) \frac{D^2G}{Dz^{\rho} Dz^{\sigma}} -
\Bigl(\frac{D\!+\!1}4\Bigr) \frac{\partial y}{\partial x^{\mu}}
\frac{\partial y}{\partial x^{\nu}} \frac{D^2G}{Dz^{\rho}
Dz^{\sigma}} \nonumber \\
& & \hspace{0cm} - H^2 \mathcal{R}_{\mu \rho} \mathcal{R}_{\nu
\sigma} \square G + \frac14 g_{\rho\sigma}(z) \frac{\partial
y}{\partial x^{\mu}} \frac{\partial y}{\partial x^{\nu}} \square G
\Biggr\} + 2 \Bigl(\frac{D\!-\!2}{D\!-\!1}\Bigr) \Biggl\{ D H^2
\mathcal{R}_{(\mu}^{~~\alpha} \mathcal{R}_{\nu) (\rho} \nonumber \\
& & \hspace{0cm} \times \frac{D^2G}{Dz^{\sigma)} Dz^{\alpha}} - H^2
g_{\rho\sigma}(z) \mathcal{R}_{(\mu}^{~~\alpha}
\mathcal{R}_{\nu)}^{~~ \beta} \frac{D^2G}{Dz^{\alpha} Dz^{\beta}}
\Biggr\} -\Biggl\{ \Bigl(\frac{D\!-\!2}{D \!-\! 1}\Bigr)
\mathcal{R}_{(\mu}^{~~\alpha} \mathcal{R}_{\nu)}^{~~\beta}
\nonumber \\
& & \hspace{0cm} \times \frac{D^4G}{Dz^{\alpha} Dz^{\beta}
Dz^{\rho} Dz^{\sigma}} - 2 \mathcal{R}_{(\mu}^{~~\alpha}
\mathcal{R}_{\nu) (\rho} \frac{D^2}{Dz^{\sigma)} Dz^{\alpha}}
\square G +\frac1{D \!-\!1} g_{\rho\sigma}(z)
\mathcal{R}_{(\mu}^{~~\alpha} \mathcal{R}_{\nu)}^{~~\beta} \nonumber \\
& & \hspace{0cm} \times \frac{D^2}{D z^{\alpha} Dz^{\beta}} \square
G \Biggr\} - \Biggl\{ \mathcal{R}_{\mu \rho} \mathcal{R}_{\nu
\sigma} \square^2 G - \frac{g_{\mu\nu}(x)}{D\!-\! 1} \Bigl[
g_{\rho\sigma}(z) \square -
\frac{D^2}{Dz^{\rho} Dx^{\sigma}} \Bigr] \square G \nonumber \\
& & \hspace{2.5cm} +\frac1{4 (D\!-\!1) H^2} \frac{\partial
y}{\partial x^{\mu}} \frac{\partial y}{\partial x^{\nu}} \Bigl[
g_{\rho\sigma}(z) \square - \frac{D^2}{Dz^{\rho} Dz^{\sigma}}
\Bigr] \square G \Biggr\} . \label{2ndP}
\end{eqnarray}

It remains just to act the derivatives in (\ref{2ndP}) and identify
the coefficients of each of the five invariant tensors from
Table~\ref{dStens}. The result of acting the d`Alembertian on an
invariant function is the same for $\square_x$ and $\square_z$,
\begin{equation}
\frac{\square_x}{H^2} \, G(y) = \frac{\square_z}{H^2} \, G(y) =
(4 y \!-\! y^2) G''(y) + D (2 \!-\! y) G'(y) \; .
\end{equation}
The other derivatives we require are,
\begin{eqnarray}
\lefteqn{\frac{\partial G}{\partial z^{\alpha}} = \frac{\partial
y}{\partial z^{\alpha}} \, G'(y) \; , \label{DID1} } \\
\lefteqn{\frac{D^2 G}{Dz^{\alpha} Dz^{\beta}} = H^2 g_{\alpha\beta}
(2 \!-\! y) G'(y) + \frac{\partial y}{\partial z^{\alpha}}
\frac{\partial y}{\partial z^{\beta}} \, G''(y) \; , \qquad
\label{DID2} } \\
\lefteqn{\frac{D^3 G}{Dz^{\alpha} Dz^{\beta} Dz^{\gamma}} = -H^2
g_{\beta\gamma} \frac{\partial y}{\partial z^{\alpha}} G'(y) + 3 H^2
g_{(\alpha\beta} \frac{\partial y}{\partial z^{\gamma)}} (2 \!-\! y)
G''(y) } \nonumber \\
& & \hspace{8cm} + \frac{\partial y}{\partial z^{\alpha}}
\frac{\partial y}{\partial z^{\beta}} \frac{\partial y}{\partial
z^{\gamma}} G'''(y) \; , \qquad \label{DID3} \\
\lefteqn{\frac{D^4 G}{Dz^{\alpha} Dz^{\beta} Dz^{\gamma}
Dz^{\delta}} = -H^4 g_{\alpha\beta} g_{\gamma\delta} (2\!-\!y) G'(y)
+ 3 H^4 g_{\alpha(\beta} g_{\gamma\delta)} (2 \!-\! y)^2 G''(y) }
\nonumber \\
& & \hspace{1.5cm} - H^2 \frac{\partial y}{\partial z^{\alpha}}
\frac{\partial y}{\partial z^{\beta}} g_{\gamma\delta} G''(y) -3 H^2
\frac{\partial y}{\partial z^{\alpha}} \frac{\partial y}{\partial
z^{(\beta}} g_{\gamma\delta)} G''(y) \nonumber \\
& & \hspace{1.5cm} + 6 H^2 g_{(\alpha\beta} \frac{\partial
y}{\partial z^{\gamma}} \frac{\partial y}{\partial z^{\delta)}} (2
\!-\! y) G'''(y) + \frac{\partial y}{\partial z^{\alpha}}
\frac{\partial y}{\partial z^{\beta}} \frac{\partial y}{\partial
z^{\gamma}} \frac{\partial y}{\partial z^{\delta}} G''''(y) \; .
\qquad \label{DID4}
\end{eqnarray}
Substituting (\ref{DID1}-\ref{DID4}) into (\ref{2ndP}) and comparing with
Table~\ref{dStens} gives,
\begin{eqnarray}
K \mathcal{C}_2^1(y) & \!\!=\!\! & -\frac14 (4y \!-\! y^2)^2 G'''' -
\frac12 (D \!+\!1) (2 \!-\! y) (4y \!-\! y^2) G''' \nonumber \\
& & \hspace{1.5cm} + \frac14 D (D \!+\! 1) (4 y \!-\! y^2) G'' -
\frac{(D \!-\! 2) D (D \!+\! 1)}{D \!-\! 1} \, G'' \; , \qquad \label{C21} \\
K \mathcal{C}_2^2(y) &\!\! =\!\! & \frac12 (2 \!-\! y) (4 y \!-\!
y^2) G'''' - (D \!+ \! 1) (4 y \!-\! y^2) G''' \nonumber \\
& & \hspace{3cm} + \frac{2 D (D \!+\! 1)}{D \!-\! 1} \, G''' -
\frac12 D (D \!+\! 1) (2 \!-\!y) G'' \; , \qquad \label{C22} \\
K \mathcal{C}_2^3(y) &\!\! =\!\! & -\Bigl( \frac{D \!-\! 2}{D \!-\!
1}\Bigr) G'''' +\frac14 (4 y \!-\! y^2) G'''' + \frac12 (D \!+\! 1)
(2 \!-\! y) G''' \nonumber \\
& & \hspace{6.5cm} - \frac14 D (D \!+\! 1) G'' \; , \qquad \label{C23} \\
K \mathcal{C}_2^4(y) & \!\!=\!\! & -\frac1{D \!-\! 1} \, (4y \!-\!
y^2) G'''' - 2 \Bigl( \frac{D \!+\! 1}{D \!-\! 1}\Bigr) (2 \!-\! y)
G''' + \frac{D (D \!+\! 1)}{D \!-\! 1} \, G'' \; , \qquad \label{C24} \\
K \mathcal{C}_2^5(y) & \!\!=\!\! & \frac1{D \!-\! 1} \, (4y \!-\!
y^2)^2 G'''' + 2 \Bigl( \frac{D \!+\! 1}{D \!-\! 1}\Bigr) (2 \!-\!
y) (4 y \!-\! y^2) G''' \nonumber \\
& & \hspace{1.5cm} - \frac{D (D \!+\! 1)}{D \!-\! 1} \, (4y \!-\!
y^2) G'' + \frac{4 (D \!-\! 2) (D \!+\! 1)}{D \!-\! 1} \, G'' \; ,
\qquad \label{C25}
\end{eqnarray}
where the constant prefactor is $K \equiv 4 (D-2)^2/(D-3)^2$.

\section{de Sitter Breaking in the Full Propagator}

\begin{table}

\vbox{\tabskip=0pt \offinterlineskip
\def\tablerule{\noalign{\hrule}}
\halign to390pt {\strut#& \vrule#\tabskip=1em plus2em& \hfil#&
\vrule#&  \hfil#\hfil& \vrule#\tabskip=0pt\cr \tablerule
\omit&height4pt&\omit&&\omit&\cr &&\omit\hidewidth k
&&\omit\hidewidth $[\mbox{}_{\mu\nu}
\mathcal{T}^k_{\rho\sigma}](x;z)$  \hidewidth&\cr
\omit&height4pt&\omit&&\omit&\cr \tablerule
\omit&height2pt&\omit&&\omit&\cr && 6 && $\frac{\partial y}{\partial
x^{(\mu}} \frac{\partial^2 y}{\partial x^{\nu)} \partial z^{(\rho}}
\frac{\partial u}{\partial z^{\sigma)}} + \frac{\partial u}{\partial
x^{(\mu}} \frac{\partial^2 y}{\partial x^{\nu)} \partial z^{(\rho}}
\frac{\partial y}{\partial z^{\sigma)}} $ &\cr
\omit&height2pt&\omit&&\omit&\cr \tablerule
\omit&height2pt&\omit&&\omit&\cr && 7 && $\frac{\partial u}{\partial
x^{(\mu}} \frac{\partial^2 y}{\partial x^{\nu)} \partial z^{(\rho}}
\frac{\partial u}{\partial z^{\sigma)}}$ &\cr
\omit&height2pt&\omit&&\omit&\cr \tablerule
\omit&height2pt&\omit&&\omit&\cr && 8 && $\frac{\partial y}{\partial
x^{\mu}} \frac{\partial y}{\partial x^{\nu}} \frac{\partial
u}{\partial z^{\rho}} \frac{\partial u}{\partial z^{\sigma}} +
\frac{\partial u}{\partial x^{\mu}} \frac{\partial u}{\partial
x^{\nu}} \frac{\partial y}{\partial z^{\rho}} \frac{\partial
y}{\partial z^{\sigma}}$ &\cr \omit&height2pt&\omit&&\omit&\cr
\tablerule \omit&height2pt&\omit&&\omit&\cr && 9 && $\frac{\partial
y}{\partial x^{(\mu}} \, \frac{\partial u }{\partial x^{\nu)}} \,
\frac{\partial y}{\partial z^{(\rho}} \, \frac{\partial u}{\partial
z^{\sigma)}}$ &\cr \omit&height2pt&\omit&&\omit&\cr \tablerule
\omit&height2pt&\omit&&\omit&\cr && 10 && $ \frac{\partial
y}{\partial x^{(\mu}} \, \frac{\partial u}{\partial x^{\nu)}}
\frac{\partial u}{
\partial z^{\rho}} \, \frac{\partial u}{\partial z^{\sigma}} +
\frac{\partial u}{\partial x^{\mu}} \, \frac{\partial u}{\partial
x^{\nu}} \, \frac{\partial y}{\partial z^{(\rho}} \, \frac{\partial
u}{\partial z^{\sigma)}}$ &\cr \omit&height2pt&\omit&&\omit&\cr
\tablerule \omit&height2pt&\omit&&\omit&\cr  && 11 && $
\frac{\partial u}{\partial x^{\mu}} \, \frac{\partial u }{\partial
x^{\nu}} \, \frac{\partial u}{\partial z^{\rho}} \, \frac{\partial
u}{\partial z^{\sigma}}$ &\cr \omit&height2pt&\omit&&\omit&\cr
\tablerule \omit&height2pt&\omit&&\omit&\cr && 12 && $H^2 [
\frac{\partial y}{\partial x^{(\mu}} \, \frac{\partial u }{\partial
x^{\nu)}} \, g_{\rho\sigma}(z) + g_{\mu\nu}(x) \, \frac{\partial
y}{\partial z^{(\rho}} \, \frac{\partial u}{\partial z^{\sigma)}}]$
&\cr \omit&height2pt&\omit&&\omit&\cr \tablerule
\omit&height2pt&\omit&&\omit&\cr && 13 && $H^2 [\frac{\partial
u}{\partial x^{\mu}} \frac{\partial u}{\partial x^{\nu}}
g_{\rho\sigma}(z) + g_{\mu\nu}(x) \frac{\partial u}{\partial
z^{\rho}} \frac{\partial u}{\partial z^{\sigma}}]$ &\cr
\omit&height2pt&\omit&&\omit&\cr \tablerule
\omit&height2pt&\omit&&\omit&\cr && 14 && $H^2 [\frac{\partial
u}{\partial x^{\mu}} \frac{\partial u}{\partial x^{\nu}}
g_{\rho\sigma}(z) - g_{\mu\nu}(x) \frac{\partial u}{\partial
z^{\rho}} \frac{\partial u}{\partial z^{\sigma}}]$ &\cr
\omit&height2pt&\omit&&\omit&\cr \tablerule}}

\caption{de Sitter breaking basis tensors for
$i[\mbox{}_{\mu\nu}\Delta_{\rho\sigma}](x;z)$. See
Table~\ref{dStens} for the five de Sitter invariant basis tensors.}

\label{nondStens}

\end{table}

In this section we act the appropriate differential projectors on
the de Sitter breaking parts of the two structure functions to
obtain explicit results. The final answer involves 14 basis tensors
multiplied by coefficient functions which depend upon $u \equiv
\ln(a_x a_z)$, $v \equiv \ln(a_x/a_z)$ and $y(x;z)$,
\begin{equation}
i \Bigl[\mbox{}_{\mu\nu} \Delta^{\rm br}_{\rho\sigma} \Bigr](x;z) =
\sum_{k=1}^{14} \Bigl[\delta \mathcal{C}^k_0(u,v,y) + \delta
\mathcal{C}^k_2(u,v,y) \Bigr] \times \Bigl[\mbox{}_{\mu\nu}
\mathcal{T}^k_{\rho\sigma}\Bigr](x;z) \; .
\end{equation}
The first five basis tensors are the invariant ones listed in
Table~\ref{dStens}. The remaining nine are noninvariant and involve
the two derivatives of $u$ which were introduced in (\ref{uders}).
These extra basis tensors are given in Table~\ref{nondStens}.

\subsection{Spin Zero Part}

The de Sitter breaking part of the spin zero contribution is,
\begin{equation}
i \Bigl[\mbox{}_{\mu\nu} \Delta^{\rm br,0}_{\rho\sigma}\Bigr](x;z) =
\mathcal{P}_{\mu\nu}(x) \times \mathcal{P}_{\rho\sigma}(z) \Bigl[
\delta S_0(u,v,y)\Bigr] \; .
\end{equation}
Recall that $\delta S_0(u,v,y)$ has the general form,
\begin{equation}
f(u,v,y) = f_1(u) + f_2(u) (y \!-\! 2) + f_3(u) \cosh(v) \; .
\end{equation}
Two useful preliminary results are,
\begin{eqnarray}
\lefteqn{\frac{D^2 f(u,v,y)}{Dx^{\mu} Dx^{\nu}} = \frac{\partial
y}{\partial x^{(\mu}} \, \frac{\partial u}{\partial x^{\nu)}} \times
2 f_2' + \frac{\partial u}{\partial x^{\mu}} \frac{\partial
u}{\partial x^{\nu}} \Biggl\{ \Bigl[-f_1' \!+\! f_1''\Bigr] }
\nonumber \\
& & \hspace{-.3cm} + \Bigl[-f_2' \!+\! f_2''\Bigr] (y \!-\! 2) +
\Bigl[f_3 \!-\! f_3' \!+\! f_3''\Bigr] \cosh(v) + \Bigl[- f_3 \!+\!
2 f_3''\Bigr] \sinh(v) \Biggr\} \nonumber \\
& & \hspace{.7cm} +H^2 g_{\mu\nu}(x) \Biggl\{ - f_1' + \Bigl[ -f_2
\!-\! f_2'\Bigr] (y \!-\! 2) - f_3' \cosh(v) - f_3 \sinh(v)\Biggr\}
, \qquad \label{d2f/dxdx} \\
\lefteqn{\frac{D^2 f(u,v,y)}{Dz^{\rho} Dz^{\sigma}} = \frac{\partial
y}{\partial z^{(\rho}} \, \frac{\partial u}{\partial z^{\sigma)}}
\times 2 f_2' + \frac{\partial u}{\partial z^{\rho}} \frac{\partial
u}{\partial z^{\sigma}} \Biggl\{ \Bigl[-f_1' \!+\! f_1''\Bigr] }
\nonumber \\
& & \hspace{-.3cm} + \Bigl[-f_2' \!+\! f_2''\Bigr] (y \!-\! 2) +
\Bigl[f_3 \!-\! f_3' \!+\! f_3''\Bigr] \cosh(v) + \Bigl[f_3 \!-\!
2 f_3''\Bigr] \sinh(v) \Biggr\} \nonumber \\
& & \hspace{.7cm} +H^2 g_{\rho\sigma}(z) \Biggl\{ - f_1' + \Bigl[
-f_2 \!-\! f_2'\Bigr] (y \!-\! 2) - f_3' \cosh(v) + f_3
\sinh(v)\Biggr\} . \qquad \label{d2f/dzdz}
\end{eqnarray}
Tracing (\ref{d2f/dxdx}) and setting it equal to the trace of
(\ref{d2f/dzdz}) implies an identity which is obeyed by all the de
Sitter breaking terms,
\begin{equation}
f_2'(u) = \frac12 \Bigl[ b_A f_3(u) \!+\! f_3'(u)\Bigr] \; .
\end{equation}

The next step is to treat the right hand side of (\ref{d2f/dxdx}) as
the source on the left hand side of (\ref{d2f/dzdz}) to infer the
result for acting four derivatives. Now employ these identities in
expression (\ref{zerodecomp}) and read off the coefficient of each
of the 14 basis tensors. The nonzero ones are,
\begin{eqnarray}
\lefteqn{\delta \mathcal{C}^5_0(u,v,y) = f_1^{W} \!-\! 2 (f_1^{MW})'
\!+\! (f_1^{S_0})'' } \nonumber \\
& & \hspace{1cm} + \Biggl[ \Bigl(f_2^{S_0} \!-\! 2 f_2^{MW} \!+\!
f_2^{W}\Bigr) \!+\! 2 \Bigl(f_2^{S_0} \!-\! f_2^{MW}\Bigr)' \!+\!
(f_2^{S_0})''\Biggr] (y \!-\! 2) \nonumber \\
& & \hspace{4cm} + \Biggl[-f_3^{S_0} \!+\! f_3^{W} \!-\! 2
(f_3^{MW})' \!+\! (f_3^{S_0})'' \Biggr] \cosh(v) \; , \qquad
\label{dC05} \\
\lefteqn{\delta \mathcal{C}^7_0(u,v,y) = 4 (f_2^{S_0})'' \; ,}
\label{dC07} \\
\lefteqn{\delta \mathcal{C}^{10}(u,v,y) = -2 (f_2^{S_0})'' + 2
(f_2^{S_0})''' \; , } \label{dC010} \\
\lefteqn{ \delta \mathcal{C}^{11}_0(u,v,y) = (f_1^{S_0})'' \!-\! 2
(f_1^{S_0})''' \!+\! (f_1^{S_0})'''' } \nonumber \\
& & \hspace{1cm} + \Biggl[ (f_2^{S_0})'' \!-\! 2 (f_2^{S_0})'''
\!+\! (f_2^{S_0})'''' \Biggr] (y \!-\! 2) \nonumber \\
& & \hspace{3cm} + \Biggl[ 2(f_3^{S_0})' \!-\! (f_3^{S_0})'' \!-\!
2 (f_3^{S_0})''' \!+\! (f_3^{S_0})'''' \Biggr] \cosh(v) \; ,
\qquad \label{dC011} \\
\lefteqn{ \delta \mathcal{C}^{12}_0(u,v,y) = 2 \Bigl( -f_2^{S_0} +
f_2^{MW}\Bigr)' - 2(f_2^{S_0})'' \; , } \label{dC012} \\
\lefteqn{ \delta \mathcal{C}^{13}(u,v,y) = -(f_1^{MW})' +
\Bigl(f_1^{S_0} \!+\! f_1^{MW}\Bigr)'' \!-\!
(f_1^{S_0})'''} \nonumber \\
& & \hspace{1cm} + \Biggl[ \Bigl(f_2^{S_0} \!-\! f_2^{MW}\Bigr)'
\!+\! (f_2^{MW})'' \!-\! (f_2^{S_0})''' \Biggr] (y \!-\! 2) \nonumber \\
& & \hspace{0cm} + \Biggl[ -f_3^{S_0} \!+\! f_3^{MW} \!+\!
\Bigl(f_3^{S_0} \!-\! f_3^{MW}\Bigr)' \!+\! \Bigl(f_3^{S_0} \!+\!
f_3^{MW} \Bigr)'' \!-\! (f_3^{S_0})''' \Biggr] \cosh(v) \; ,
\qquad \label{dC013} \\
\lefteqn{\delta \mathcal{C}^{14}_0(u,v,y) = \Biggl[ f_3^{S_0} \!-\!
f_3^{MW} \!+\! 2 (f_3^{MW})' \!-\! (f_3^{S_0})'' \Biggr] \sinh(v) \;
. } \label{dC014}
\end{eqnarray}

The final step is to substitute the Section 3.1 results (\ref{fW}),
(\ref{fMW1}-\ref{fMW3}) and (\ref{fS01}-\ref{fS03}) into relations
(\ref{dC05}-\ref{dC014}). One surprising consequence is that all the
de Sitter breaking contributions from the $W$-type propagator cancel
out. These are homogeneous solutions of the spin zero propagator
equation. The nonzero contributions derive from the $M$-type
propagator. For the values $k=5$, $k=11$ and $k=13$ the coefficients
take the form,
\begin{eqnarray}
\lefteqn{ \delta \mathcal{C}^k_0(u,v,y) = \frac{k_M}{H^4} \Biggl\{
\Bigl[ (u \!+\! C_M) A^k_1 \!+\! B^k_1\Bigr] \times e^{(b_M - b_A)
u} + \Bigl[ (u \!+\! C_M) A^k_2 \!+\! B^k_2\Bigr] } \nonumber \\
& & \hspace{0cm} \times e^{(b_M - b_W) u} \times (y \!-\! 2) +
\Bigl[ (u \!+\! C_M) A^k_3 \!+\! B^k_3\Bigr] \times e^{(b_M - b_W)
u} \times \cosh(v) \Biggr\} . \qquad \label{masterform}
\end{eqnarray}
For $k=7$, $k=10$ and $k=12$ only the $f_2$ terms contribute,
\begin{equation}
\delta \mathcal{C}^k_0(u,v,y) = \frac{k_M}{H^4} \Bigl[ (u \!+\! C_M)
A^k_2 \!+\! B^k_2\Bigr] \times e^{(b_M - b_W) u} \; .
\label{submaster1}
\end{equation}
And the final coefficient takes the form,
\begin{equation}
\delta \mathcal{C}^{14}_0(u,v,y) = \frac{k_M}{H^4} \Bigl[ (u \!+\!
C_M) A^k_3 \!+\! B^k_3\Bigr] \times e^{(b_M - b_W) u} \times
\sinh(v) \; . \label{submaster2}
\end{equation}
Tables~\ref{spin0AB1}-\ref{spin0AB2} list the constants $A^k_{1-3}$
and $B^k_{1-3}$ for each of the nonzero coefficients.

\begin{table}

\vbox{\tabskip=0pt \offinterlineskip
\def\tablerule{\noalign{\hrule}}
\halign to390pt {\strut#& \vrule#\tabskip=1em plus2em& \hfil#&
\vrule#&  \hfil#\hfil& \vrule#\tabskip=0pt\cr \tablerule
\omit&height4pt&\omit&&\omit&\cr &&\omit\hidewidth
\textrm{Constants} \hidewidth&&\omit\hidewidth \textrm{Values}
\hidewidth&\cr \omit&height4pt&\omit&&\omit&\cr \tablerule
\omit&height2pt&\omit&&\omit&\cr && $A^5_1$ && $\frac{b_A -
b_M}{2b_{A} b_M}$ &\cr \omit&height2pt&\omit&&\omit&\cr \tablerule
\omit&height2pt&\omit&&\omit&\cr && $B^{5}_1$ && $\frac{1 + 6 b_A +
2 b_A^2 + (-4 - 2 b_A) b_M} {2(D-2)b_{A}b_{M}}$ &\cr
\omit&height2pt&\omit&&\omit&\cr \tablerule
\omit&height2pt&\omit&&\omit&\cr && $A^5_2$ && $\frac{4 b_A + 2
b_A^2 - 2 b_A b_M} {8 b_A^2 b_M (b_M-b_W)}$ &\cr
\omit&height2pt&\omit&&\omit&\cr \tablerule
\omit&height2pt&\omit&&\omit&\cr && $B^5_2$ && $\frac{-22 b_A - 74
b_A^2 - 48 b_A^3 -
 8 b_A^4 + (2 + 26 b_A + 32 b_A^2 + 8 b_A^3) b_M}
 {8 (D-2) b_A^2 b_M (b_M-b_W)^2}$ &\cr
\omit&height2pt&\omit&&\omit&\cr \tablerule
\omit&height2pt&\omit&&\omit&\cr && $A^5_3$ && $\frac{6 b_A + 2
b_A^2 + (-2 - 2 b_A) b_M} {4 b_A^2 b_M (b_M-1)}$ &\cr
\omit&height2pt&\omit&&\omit&\cr \tablerule
\omit&height2pt&\omit&&\omit&\cr  && $B^5_3$ && $\frac{-2 - 48 b_A -
84 b_A^2 - 36 b_A^3 -
 4 b_A^4 + (6 + 44 b_A + 28 b_A^2 + 4 b_A^3) b_M}
 {4 (D-2) b_A^2 b_M (b_M-1)^2}$ &\cr
\omit&height2pt&\omit&&\omit&\cr \tablerule
\omit&height2pt&\omit&&\omit&\cr && $A^7_2$ && $\frac{-1 - b_A +
b_M}{2 b_A^2 b_M}$ &\cr \omit&height2pt&\omit&&\omit&\cr \tablerule
\omit&height2pt&\omit&&\omit&\cr && $B^7_2$ && $\frac{-1 - 6 b_A - 2
b_A^2 + (2 + 2 b_A) b_M} {2 (D-2) b_A^2 b_M }$ &\cr
\omit&height2pt&\omit&&\omit&\cr \tablerule
\omit&height2pt&\omit&&\omit&\cr && $A^{10}_2$ && $\frac{2 + 7 b_A +
2 b_A^2 + (-3 - 2 b_A) b_M} {4 b_A^2 b_M}$ &\cr
\omit&height2pt&\omit&&\omit&\cr \tablerule
\omit&height2pt&\omit&&\omit&\cr && $B^{10}_2$ && $\frac{3 + 20 b_A
+ 18 b_A^2 + 4 b_A^3 + (-6 - 10 b_A - 4 b_A^2) b_M}{4 (D-2) b_A^2
b_M}$ &\cr \omit&height2pt&\omit&&\omit&\cr \tablerule
\omit&height2pt&\omit&&\omit&\cr && $A^{11}_1$ && $\frac{9 b_A + 16
b_A^2 + 4 b_A^3 + (-1 - 8 b_A - 4 b_A^2) b_M}{2 b_A b_M}$ &\cr
\omit&height2pt&\omit&&\omit&\cr \tablerule
\omit&height2pt&\omit&&\omit&\cr && $B^{11}_1$ && $\frac{1 + 22 b_A
+ 40 b_A^2 + 36 b_A^3 +
 8 b_A^4 + (-4 - 16 b_A - 20 b_A^2 - 8 b_A^3) b_M}{2 (D-2) b_A b_M }$ &\cr
\omit&height2pt&\omit&&\omit&\cr \tablerule
\omit&height2pt&\omit&&\omit&\cr && $A^{11}_2$ && $\frac{-4 - 28 b_A
- 22 b_A^2 - 4 b_A^3 + (8 + 14 b_A + 4 b_A^2) b_M}{8 b_A^2 b_M}$
&\cr \omit&height2pt&\omit&&\omit&\cr \tablerule
\omit&height2pt&\omit&&\omit&\cr && $B^{11}_2$ && $\frac{-8 - 70 b_A
- 90 b_A^2 - 48 b_A^3 -
 8 b_A^4 + (18 + 42 b_A + 32 b_A^2 + 8 b_A^3) b_M}{8 (D-2) b_A^2 b_M}$ &\cr
\omit&height2pt&\omit&&\omit&\cr \tablerule
\omit&height2pt&\omit&&\omit&\cr && $A^{11}_3$ && $\frac{50 b_A +
110 b_A^2 + 56 b_A^3 +
 8 b_A^4 + (-6 - 46 b_A - 40 b_A^2 - 8 b_A^3) b_M}{4 b_A^2 b_M (b_M-1)}$ &\cr
\omit&height2pt&\omit&&\omit&\cr \tablerule
\omit&height2pt&\omit&&\omit&\cr && $B^{11}_3$ && $\frac{-6 - 168
b_A - 720 b_A^2 - 1060 b_A^3 - 668 b_A^4 - 176 b_A^5 -16 b_A^6}{4
(D-2) b_A^2 b_M (b_M-1)^2}$ &\cr \omit&height2pt&\omit&&\omit&\cr &&
\omit && $+ \frac{(22 + 204 b_A + 460 b_A^2 + 412 b_A^3 + 144 b_A^4
+ 16 b_A^5) b_M}{4 (D-2) b_A^2 b_M (b_M-1)^2}$ &\cr
\omit&height2pt&\omit&&\omit&\cr \tablerule
\omit&height2pt&\omit&&\omit&\cr && $A^{12}_2$ && $\frac{b_A -
b_M}{4 b_A^2 b_M}$ &\cr \omit&height2pt&\omit&&\omit&\cr \tablerule
\omit&height2pt&\omit&&\omit&\cr && $B^{12}_2$ && $\frac{1 + 6 b_A +
2 b_A^2 + (-2 - 2 b_A) b_M} {4 (D-2) b_A^2 b_M}$ &\cr
\omit&height2pt&\omit&&\omit&\cr \tablerule}}

\caption{The nonzero constants of $A^k_{1-3}$ and $B^k_{1-3}$ in
expressions (\ref{masterform}-\ref{submaster2}) for $\delta
\mathcal{C}_0^k(u,v,y)$.}

\label{spin0AB1}

\end{table}

\begin{table}

\vbox{\tabskip=0pt \offinterlineskip
\def\tablerule{\noalign{\hrule}}
\halign to390pt {\strut#& \vrule#\tabskip=1em plus2em& \hfil#&
\vrule#&  \hfil#\hfil& \vrule#\tabskip=0pt\cr \tablerule
\omit&height4pt&\omit&&\omit&\cr &&\omit\hidewidth
\textrm{Constants} \hidewidth&&\omit\hidewidth \textrm{Values}
\hidewidth&\cr \omit&height4pt&\omit&&\omit&\cr \tablerule
\omit&height2pt&\omit&&\omit&\cr && $A^{13}_1$ && $\frac{5 b_A + 2
b_A^2 + (-1 - 2 b_A) b_M}{2 b_A b_M}$ &\cr
\omit&height2pt&\omit&&\omit&\cr \tablerule
\omit&height2pt&\omit&&\omit&\cr && $B^{13}_1$ && $\frac{1 + 16 b_A
+ 16 b_A^2 + 4 b_A^3 + (-4 - 8 b_A - 4 b_A^2) b_M} {2 (D - 2) b_A
b_M}$ &\cr \omit&height2pt&\omit&&\omit&\cr \tablerule
\omit&height2pt&\omit&&\omit&\cr && $A^{13}_2$ && $\frac{-6 b_A-2
b_A^2+(2+2 b_A) b_M}{8 b_A^2 b_M}$ &\cr
\omit&height2pt&\omit&&\omit&\cr \tablerule
\omit&height2pt&\omit&&\omit&\cr && $B^{13}_2$ && $\frac{-1-11 b_A-9
b_A^2-2 b_A^3+(3+5 b_A+2 b_A^2) b_M} {4 (D-2) b_A^2 b_M}$ &\cr
\omit&height2pt&\omit&&\omit&\cr \tablerule
\omit&height2pt&\omit&&\omit&\cr && $A^{13}_3$ && $\frac{20 b_A+20
b_A^2+4 b_A^3+(-4-12 b_A-4 b_A^2) b_M}{4 b_A^2 b_M (b_M-1)}$ &\cr
\omit&height2pt&\omit&&\omit&\cr \tablerule
\omit&height2pt&\omit&&\omit&\cr && $B^{13}_3$ && $\frac{-4-104
b_A-302 b_A^2-256 b_A^3-80 b_A^4-8 b_A^5 +(14+108 b_A+144 b_A^2+64
b_A^3+8 b_A^4) b_M}{4 (D-2) b_A^2 b_M (b_M-1)^2}$ &\cr
\omit&height2pt&\omit&&\omit&\cr \tablerule
\omit&height2pt&\omit&&\omit&\cr && $A^{14}_3$ && $\frac{-6 b_A - 2
b_A^2 + (2 + 2 b_A) b_M}{4 b_A^2 b_M (b_M-1)}$ &\cr
\omit&height2pt&\omit&&\omit&\cr \tablerule
\omit&height2pt&\omit&&\omit&\cr && $B^{14}_3$ && $\frac{2 + 56 b_A
+ 86 b_A^2 + 36 b_A^3 +
 4 b_A^4 + (-8 - 44 b_A - 28 b_A^2 - 4 b_A^3) b_M}
 {4 (D-2) b_A^2 b_M (b_M-1)^2}$ &\cr
\omit&height2pt&\omit&&\omit&\cr \tablerule}}

\caption{The nonzero constants of $A^k_{1-3}$ and $B^k_{1-3}$ in
expressions (\ref{masterform}-\ref{submaster2}) for $\delta
\mathcal{C}_0^k(u,v,y)$.}

\label{spin0AB2}

\end{table}

\subsection{Spin Two Part}

The de Sitter breaking contribution from the spin two part is,
\begin{equation}
i\Bigl[\mbox{}_{\mu\nu} \Delta^{\rm br,2}_{\rho\sigma}\Bigr](x;z) =
\frac1{4 H^4} \mathbf{P}_{\mu\nu}^{~~\alpha\beta}(x) \times
\mathbf{P}_{\rho\sigma}^{~~\kappa\lambda}(z) \Bigl[
\mathcal{R}_{\alpha\kappa}(x;z) \mathcal{R}_{\beta\lambda}(x;z)
\delta S_2(u)\Bigr] \; .
\end{equation}
In acting the first projector
$\mathbf{P}_{\mu\nu}^{~~\alpha\beta}(x)$ we can use the result
(\ref{1stP}) from the previous section, which is valid without
regard to the spacetime dependence of the structure function.
Because the additional simplification (\ref{1stPsimp}) only applies
for de Sitter invariant functions of $y$, we instead express
(\ref{1stP}) in the form,
\begin{eqnarray}
\lefteqn{ 2\Bigl(\frac{D \!-\! 2}{D \!-\! 3}\Bigr)
\mathbf{P}_{\mu\nu}^{~~\alpha\beta}(x) \Bigl[ \mathcal{R}_{\alpha
\kappa} \mathcal{R}_{\beta \lambda} F \Bigr] = {\rm Trace\ Terms} }
\nonumber \\
& & + \mathcal{R}_{\mu \kappa} \mathcal{R}_{\nu \lambda} \times G_A
+ \mathcal{R}^{\alpha}_{~\kappa} \mathcal{R}_{\alpha \lambda} \times
(G_B)_{\mu\nu} + \mathcal{R}_{\mu (\kappa}
\mathcal{R}^{\alpha}_{~\lambda)} \times (G_C)_{\alpha \nu} \nonumber \\
& & + \mathcal{R}^{\alpha}_{~ \kappa} \mathcal{R}^{\beta}_{~\lambda}
\times (G_D)_{\alpha\beta\mu\nu} + \mathcal{R}_{\mu (\kappa}
\frac{\partial y}{\partial z^{\lambda)}} \times (G_E)_{\nu} +
\mathcal{R}^{\alpha}_{~ (\kappa} \frac{\partial y}{\partial
z^{\lambda)}} \times (G_F)_{\mu\nu\alpha} \; . \qquad \label{newP}
\end{eqnarray}
To save space, the right hand side of (\ref{newP}) is understood to
be symmetrized on $\mu$ and $\nu$, which is relevant to the 3rd and
5th terms. The six $G_I$'s are,
\begin{eqnarray}
G_A & = & -\square_x^2 F -\frac{(D \!-\! 2) (D \!+\! 1)}{D \!-\! 1}
H^2 \square_x F \; , \qquad \\
(G_B)_{\mu\nu} & = & \frac{g_{\mu\nu}}{D \!-\! 1} \square_x^2 F -
\frac1{D \!-\! 1} D_{\mu} D_{\nu} \square_x F \nonumber \\
& & \hspace{-.5cm} - \frac{(D \!-\!2) (D \!+\! 1)}{D \!-\! 1} H^2
g_{\mu\nu} \square_x F + \frac{(D \!-\! 2)(D \!+\! 1)^2}{D \!-\! 1}
H^2 D_{\mu} D_{\nu} F \; , \qquad \\
(G_C)_{\nu\alpha} & = & 2 D_{\nu} D_{\alpha} \square_x F + \frac{2
(D \!-\! 2) D}{D \!-\! 1} H^2 D_{\nu} D_{\alpha} F \; , \qquad \\
(G_D)_{\alpha\beta\mu\nu} & = & - \frac{g_{\mu\nu}}{D \!-\! 1}
D_{\alpha} D_{\beta} \square F -\Bigl(\frac{D \!-\! 2}{D \!-\!
1}\Bigr) D_{\alpha} D_{\beta} D_{\mu} D_{\nu} F \nonumber \\
& & \hspace{5cm} -2 \Bigl(\frac{D \!-\! 2}{D \!-\! 1}\Bigr) H^2
g_{\mu\nu} D_{\alpha} D_{\beta} F \; , \qquad \\
(G_E)_{\nu} & = & \frac{(D \!-\! 2) (D\!+\! 1)}{D \!-\! 1}
D_{\nu} \square_x F + (D\!-\!2) (D\!+\! 1) H^2 D_{\nu} F \; , \\
(G_F)_{\mu\nu\alpha} & = & - \frac{(D\!-\!2)(D\!+\!1)}{D \!-\! 1}
D_{(\mu} D_{\nu)} D_{\alpha} F \; . \qquad
\end{eqnarray}

The first four terms of (\ref{newP}) involve two factors of
$\mathcal{R}$. The result of acting $\mathbf{P}_{\rho\sigma}^{~~
\kappa\lambda}(z)$ on these terms can be read off expression
(\ref{1stPsimp}) by merely interchanging $x^{\mu}$ with $z^{\mu}$
and their respective index groups. The final two terms of
(\ref{newP}) involve a factor of $\mathcal{R}$ times one of
$\partial y/\partial z$. Acting $\mathbf{P}_{\rho\sigma}^{~~
\kappa\lambda}(z)$ on these terms requires some new identities such
as,
\begin{equation}
D_{\rho} \Bigl[ \mathcal{R}_{\mu \kappa} \frac{\partial y}{\partial
z^{\lambda}} G\Bigr] = \frac12 g_{\rho\kappa} \frac{\partial
y}{\partial x^{\mu}} \frac{\partial y}{\partial z^{\lambda}} G + H^2
(2 \!-\! y) g_{\rho \lambda} \mathcal{R}_{\mu \kappa} G +
\mathcal{R}_{\mu \kappa} \frac{\partial y}{\partial z^{\lambda}}
\frac{D G}{D z^{\rho}} \; .
\end{equation}
The Appendix gives 7 related identities which allow us to derive,
\begin{eqnarray}
\lefteqn{ \hspace{-.1cm} 2 \Bigl( \frac{D \!-\! 2}{D
\!-\! 3}\Bigr) \mathbf{P}_{\rho\sigma}^{~~ \kappa\lambda}(z) \Bigl[
\mathcal{R}_{\mu\kappa} \frac{\partial y}{\partial z^{\lambda}} G
\Bigr] = } \nonumber \\
& & \hspace{-.7cm} \Bigl[-\delta^{\kappa}_{(\rho}
\delta^{\lambda}_{\sigma)} g^{\theta\phi} \!+\! \frac{g_{\rho\sigma}
g^{\kappa\lambda} g^{\theta\phi}}{D \!-\! 1} \!-\!
\frac{\delta^{\theta}_{(\rho} \delta^{\phi}_{\sigma)}
g^{\kappa\lambda}}{ D\!-\! 1} \Bigr] \mathcal{R}_{\mu\kappa}
\frac{\partial y}{\partial z^{\lambda}}
D_{\theta} D_{\phi} \square_z G \nonumber \\
& & \hspace{-.7cm} - \frac{(D \!-\!2) (D\!+\! 1)}{D \!-\! 1}
\Bigl[\delta^{\kappa}_{(\rho} \delta^{\lambda}_{\sigma)}
g^{\theta\phi} \!\!+\! g_{\rho\sigma} g^{\kappa\lambda}
g^{\theta\phi} \!\!-\! (D \!+\!1) \delta^{\theta}_{(\rho}
\delta^{\phi}_{\sigma)} g^{\kappa\lambda} \Bigr]
\mathcal{R}_{\mu\kappa} \frac{\partial
y}{\partial z^{\lambda}} H^2 D_{\theta} D_{\phi} G \nonumber \\
& & \hspace{-.7cm} + \frac{(D\!-\!2)(D\!+\!1)}{D \!-\! 1} \Bigl[
\delta^{\kappa}_{(\rho} \delta^{\lambda}_{\sigma)} g^{\theta\phi}
\!-\! \delta^{\theta}_{\rho} \delta^{\phi}_{\sigma}
g^{\kappa\lambda} \Bigr] \Biggl\{ \frac12 \frac{\partial y}{\partial
x^{\mu}} \frac{\partial y}{\partial z^{\kappa}} \!+\! (2 \!-\!y)
H^2 \mathcal{R}_{\mu \kappa} \Biggr\} D_{\lambda} D_{\theta} D_{\phi} G
\nonumber \\
& & \hspace{-.7cm} + \frac{(D\!-\!2)(D\!+\!1)}{D \!-\! 1} \Bigl[ D
\delta^{\kappa}_{(\rho} \delta^{\lambda}_{\sigma)} \!-\!
g_{\rho\sigma} g^{\kappa\lambda} \Bigr] \Biggl\{ \frac12
\frac{\partial y}{\partial x^{\mu}} \frac{\partial y}{\partial
z^{\kappa}} \!+\! (2 \!-\!y) H^2
\mathcal{R}_{\mu \kappa} \Biggr\} H^2 D_{\lambda} G \nonumber \\
& & \hspace{-.7cm} +  \Bigl[ 2 g^{\theta (\kappa}
\delta^{\lambda)}_{(\rho} \delta^{\phi}_{\sigma)} g^{\psi \chi}
\!-\! \frac{g_{\rho\sigma} g^{\theta \kappa} g^{\phi\lambda}
g^{\psi\chi}}{D \!-\! 1} \!-\! \Bigl( \frac{D \!-\! 2}{D \!-\!
1}\Bigr) \delta^{\psi}_{(\rho} \delta^{\chi}_{\sigma)} g^{\theta
(\kappa} g^{\lambda) \phi} \Bigr] \mathcal{R}_{\mu\kappa}
\frac{\partial y}{\partial z^{\lambda}} D_{\theta} D_{\phi}
\nonumber \\
& & \hspace{0cm} \times D_{\psi} D_{\chi} G + 2 \Bigl( \frac{D \!-\!
2}{D \!-\! 1}\Bigr) \Bigl[ D g^{\theta (\kappa}
\delta^{\lambda)}_{(\rho} \delta^{\phi}_{\sigma)} \!-\!
g_{\rho\sigma} g^{\theta\kappa} g^{\phi\lambda}\Bigr]
\mathcal{R}_{\mu\kappa} \frac{\partial y}{\partial z^{\lambda}} H^2
D_{\theta} D_{\phi} G \; . \qquad \label{newP2}
\end{eqnarray}

The next step is to act the various derivatives of $F(u) = \delta
S_2(u)/4 H^4$. There is always at least one derivative with respect
to each coordinate, so the simplest case is,
\begin{equation}
D^x_{\alpha} D^z_{\kappa} F = u_{\alpha} u_{\kappa} F'' \; .
\end{equation}
There can be up two four derivatives with respect to each
coordinate, making for ten cases. These are given in the Appendix.

After acting the derivatives one makes use of the tensor contraction
identities of section 2.1 and extracts the coefficient of each of
the basis tensors from Table~\ref{dStens} and Table~\ref{nondStens}.
Because the intermediate expressions become quite lengthy, the
differentiation and tensor operations were performed using the
symbolic manipulation program Mathematica. The final result is that
the nonzero coefficient functions are all proportional to $\delta
\mathcal{C}^1_2$,
\begin{eqnarray}
\delta \mathcal{C}^5_2 = -\frac{4}{D \!-\! 1} \times \delta
\mathcal{C}^1_2 & , & \delta \mathcal{C}^9_2 = +2 \times \delta
\mathcal{C}^1_2 \; , \label{dC52} \\
\delta \mathcal{C}^6_2 = -2 \times \delta \mathcal{C}^1_2 & , &
\delta \mathcal{C}^{10}_2 = +2 (2 \!-\! y) \times \delta
\mathcal{C}^1_2 \; , \label{dC62} \\
\delta \mathcal{C}^7_2 = -2 (2 \!-\! y) \times \delta
\mathcal{C}^1_2 & , & \delta \mathcal{C}^{11}_2 = \Bigl[-\frac4{D
\!-\! 1} + (2\!-\!y)^2\Bigr] \times \delta\mathcal{C}^1_2 \; ,
\label{dC72} \\
\delta \mathcal{C}^8_2 = +1 \times \delta \mathcal{C}^1_2 & , &
\delta \mathcal{C}^{13}_2 = -\frac4{D \!-\! 1} \times \delta
\mathcal{C}^1_2 \; . \label{dC82}
\end{eqnarray}
The coefficient function $\delta \mathcal{C}^1_2$ involves
derivatives of the Sitter breaking part of the spin two structure
function $\delta S_2(u)$, given in (\ref{deltaS2}),
\begin{eqnarray}
\delta \mathcal{C}^1_2 & = & \frac1{4 H^4} \times \frac14 \Bigl(
\frac{D \!-\! 3}{D \!-\! 2}\Bigr)^2 \times \frac14 (D\!-\!2)^2
(D\!-\! 1)^2 H^4 \nonumber \\
& & \hspace{3cm} \times \Biggl\{ \delta S_2''(u) - \Bigl[2 \!-\!
\frac2{D \!-\! 2} \!-\! \frac2{D \!-\! 1} \Bigr] \delta S_2'''(u)
\Biggr\} , \qquad \\
& = & \frac{k}{2 H^4} \Biggl\{ \ln(4 a_x a_z) \!+\! 2 \psi\Bigl(
\frac{D \!-\! 1}2\Bigr) \!-\! 4 \!+\! \frac1{D \!-\! 1} \Biggr\} .
\label{dC12}
\end{eqnarray}
Recall that the constant $k$ was defined in (\ref{kdefs}).

Great simplifications arise when the various spin 2 de Sitter
breaking terms are combined. For example, from Tables~\ref{dStens}
and \ref{nondStens} we see that the terms proportional to $-4/(D-1)$
sum up to give,
\begin{equation}
-\frac{4 \delta \mathcal{C}^1_2}{D \!-\! 1} \Biggl\{ \Bigl[
\mbox{}_{\mu\nu} \mathcal{T}^5_{\rho\sigma}\Bigr] + \Bigl[
\mbox{}_{\mu\nu}\mathcal{T}^{11}_{\rho\sigma}\Bigr] + \Bigl[
\mbox{}_{\mu\nu} \mathcal{T}^{13}_{\rho\sigma}\Bigr] \Biggr\} = 2
H^4 \delta \mathcal{C}^1_2 \times - \frac{2}{D \!-\!1}
g^{\perp}_{\mu\nu}(x) g^{\perp}_{\rho\sigma}(z) \; . \label{part1}
\end{equation}
Here and henceforth we define the purely spatial, tangent space
metric,
\begin{equation}
g^{\perp}_{\mu\nu}(x) \equiv g_{\mu\nu}(x) + \frac1{H^2}
\frac{\partial u}{\partial x^{\mu}} \frac{\partial u}{\partial
x^{\nu}} = a_x^2 \Bigl[ \eta_{\mu\nu} + \delta^0_{\mu}
\delta^0_{\nu}\Bigr] \; . \label{gperp}
\end{equation}
An even greater simplification attends the remaining terms,
\begin{eqnarray}
\lefteqn{ \delta \mathcal{C}^1_2 \Biggl\{ \Bigl[ \mbox{}_{\mu\nu}
\mathcal{T}^1_{\rho\sigma}\Bigr] - 2 \Bigl[ \mbox{}_{\mu\nu}
\mathcal{T}^6_{\rho\sigma}\Bigr] - 2 (2 \!-\!y) \Bigl[
\mbox{}_{\mu\nu} \mathcal{T}^7_{\rho\sigma}\Bigr] } \nonumber \\
& & \hspace{1cm} + \Bigl[ \mbox{}_{\mu\nu}
\mathcal{T}^8_{\rho\sigma}\Bigr] + 2 \Bigl[ \mbox{}_{\mu\nu}
\mathcal{T}^9_{\rho\sigma}\Bigr] + 2(2 \!-\! y) \Bigl[
\mbox{}_{\mu\nu} \mathcal{T}^{10}_{\rho\sigma}\Bigr] + (2 \!-\! y)^2
\Bigl[ \mbox{}_{\mu\nu} \mathcal{T}^{11}_{\rho\sigma}\Bigr] \Biggr\}
, \qquad \\
& & = 2 H^4 \delta \mathcal{C}^1_2 \times \Bigl[
\mathcal{R}^{\perp}_{\mu\rho} \mathcal{R}^{\perp}_{\nu\sigma} +
\mathcal{R}^{\perp}_{\mu\sigma} \mathcal{R}^{\perp}_{\nu\rho} \Bigr]
\; . \qquad \label{part2}
\end{eqnarray}
The purely spatial version of $\mathcal{R}$ is,
\begin{eqnarray}
\mathcal{R}^{\perp}_{\mu\rho}(x;z) & \equiv & -\frac1{2 H^2}
\Biggl\{ \frac{\partial^2 y}{\partial x^{\mu} \partial z^{\rho}}
\!-\! \frac{\partial y}{\partial x^{\mu}} \frac{\partial u}{\partial
z^{\rho}} \!-\! \frac{\partial u}{\partial x^{\mu}} \frac{\partial
y}{\partial z^{\rho}} - (2 \!-\! y) \frac{\partial u}{\partial
x^{\mu}} \frac{\partial u}{\partial z^{\rho}} \Biggr\} \; , \qquad
\\
& = & a_x a_z \Bigl[ \eta_{\mu\rho} + \delta^0_{\mu}
\delta^0_{\rho} \Bigr] \; . \label{Rperp}
\end{eqnarray}
Combining expressions (\ref{dC12}), (\ref{part1}) and (\ref{part2})
gives our final form for the spin 2 de Sitter breaking part,
\begin{eqnarray}
\lefteqn{ i\Bigl[\mbox{}_{\mu\nu} \Delta^{\rm
br,2}_{\rho\sigma}\Bigr](x;z) = k \Biggl[ \ln(4 a_x a_z) \!+\! 2
\psi\Bigl( \frac{D \!-\!1}2\Bigr) \!-\! 4 \!+\! \frac1{D \!-\!
1}\Biggr] } \nonumber \\
& & \hspace{5cm} \times \Biggl\{ \mathcal{R}^{\perp}_{\mu\rho}
\mathcal{R}^{\perp}_{\nu\sigma} + \mathcal{R}^{\perp}_{\mu\sigma}
\mathcal{R}^{\perp}_{\nu\sigma} - \frac2{D \!-\! 1} \,
g^{\perp}_{\mu\nu} g^{\perp}_{\rho\sigma} \Biggr\} . \qquad
\label{nondS2}
\end{eqnarray}

A few comments about (\ref{nondS2}) derive from simple properties of
the tangent space tensors $g^{\perp}_{\mu\nu}$ and
$\mathcal{R}^{\perp}_{\mu\nu}$. First, note the trace identities,
\begin{equation}
g^{\perp}_{\alpha\mu} \times g^{\alpha\beta} g^{\mu\nu} \times
g^{\perp}_{\beta\nu} = D \!-\! 1 \qquad , \qquad
\mathcal{R}^{\perp}_{\mu\rho} \times g^{\mu\nu} \times
\mathcal{R}^{\perp}_{\nu\sigma} = g^{\perp}_{\rho\sigma} \; .
\label{cIDs}
\end{equation}
These relations imply that (\ref{nondS2}) is traceless. Covariant
differentiation with respect to $x^{\alpha}$ yields,
\begin{equation}
D_{\alpha} g^{\perp}_{\mu\nu} = -2 g^{\perp}_{\alpha (\mu}
\frac{\partial u}{\partial x^{\nu)}} \qquad , \qquad D_{\alpha}
\mathcal{R}^{\perp}_{\mu\rho} = -\frac{\partial u}{\partial x^{\mu}}
\mathcal{R}^{\perp}_{\alpha\rho} \; . \label{dIDs}
\end{equation}
(Of course similar results apply to differentiation with respect to
$z^{\kappa}$.) Together with (\ref{cIDs}), and the orthogonality
with respect to derivatives of $u$, relation (\ref{dIDs}) implies
that (\ref{nondS2}) is transverse.

Finally, it is interesting to compare (\ref{nondS2}) with the
infrared logarithm part of the only propagator so far used to make
loop computations \cite{RPW,TW2}. That result was derived in a
noncovariant, average gauge which is not subject to the topological
obstacle \cite{MTW2} because it cannot be extended beyond the open
coordinate submanifold. When the multiplied by $a_x^2 a_z^2$ to
account for conformally rescaling the graviton field the infrared
logarithm part of that propagator is \cite{RPW,TW2},
\begin{equation}
k \ln(a_x a_z) \Biggl\{ \mathcal{R}^{\perp}_{\mu\rho}
\mathcal{R}^{\perp}_{\nu\sigma} + \mathcal{R}^{\perp}_{\mu\sigma}
\mathcal{R}^{\perp}_{\nu\sigma} - \frac2{D \!-\! 3} \,
g^{\perp}_{\mu\nu} g^{\perp}_{\rho\sigma} \Biggr\} \; .
\label{oldprop}
\end{equation}
Note that the mixed index ($\mathcal{R}_{\mu\rho}
\mathcal{R}_{\nu\sigma} + \mathcal{R}_{\mu\sigma}
\mathcal{R}_{\nu\rho}$) parts of (\ref{oldprop}) agree precisely
with those of (\ref{nondS2}). The unmixed part of (\ref{oldprop})
includes contributions from the gauge-dependent, spin 0 part so
there is no reason they should agree in different gauges. But the
spin 2 parts should be gauge independent, and they do seem to be.

\section{The Coincidence Limit}

``Coincidence'' means taking the coordinate $z^{\mu}$ equal to
$x^{\mu}$ in $i[\mbox{}_{\mu\nu}\Delta_{\rho\sigma}](x;z)$. Table
\ref{tenscoin} gives the coincidence limit of each of the 14 basis
tensors. The de Sitter length function $y(x,z)$ vanishes at
coincidence. The rules of dimensional regularization imply that all
$D$-dependent powers of $y$ vanish as well \cite{OW}. Coincidence
also makes the variable $v = \ln(\frac{a_x}{a_z})$ go to zero, which
implies $\cosh(v) \rightarrow 1$ and $\sinh(v) \rightarrow 0$. The
variable $u = \ln(a_x a_z)$ goes to $2 \ln(a)$ at coincidence.

\begin{table}

\vbox{\tabskip=0pt \offinterlineskip
\def\tablerule{\noalign{\hrule}}
\halign to390pt {\strut#& \vrule#\tabskip=1em plus2em& \hfil#&
\vrule#& \hfil#\hfil& \vrule#& \hfil#& \vrule#& \hfil#\hfil&
\vrule#\tabskip=0pt\cr \tablerule
\omit&height4pt&\omit&&\omit&&\omit&&\omit&\cr &&\omit\hidewidth k
&&\omit\hidewidth $[\mbox{}_{\mu\nu}
\mathcal{T}^k_{\rho\sigma}](x;x)$ \hidewidth&& \omit\hidewidth k
\hidewidth&& \omit\hidewidth$[\mbox{}_{\mu\nu}
\mathcal{T}^k_{\rho\sigma}](x;x)$ \hidewidth&\cr
\omit&height4pt&\omit&&\omit&&\omit&&\omit&\cr \tablerule
\omit&height2pt&\omit&&\omit&&\omit&&\omit&\cr && 1 && $4 H^4 g_{\mu
(\rho} g_{\sigma) \nu}$ && 8 && $0$ &\cr
\omit&height2pt&\omit&&\omit&&\omit&&\omit&\cr \tablerule
\omit&height2pt&\omit&&\omit&&\omit&&\omit&\cr && 2 && $0$ && 9 &&
$0$ &\cr \omit&height2pt&\omit&&\omit&&\omit&&\omit&\cr \tablerule
\omit&height2pt&\omit&&\omit&&\omit&&\omit&\cr && 3 && $0$ && 10 &&
$0$ &\cr \omit&height2pt&\omit&&\omit&&\omit&&\omit&\cr \tablerule
\omit&height2pt&\omit&&\omit&&\omit&&\omit&\cr && 4 && $0$ && 11 &&
$H^4 a^4 \delta^0_{\mu} \delta^0_{\nu} \delta^0_{\rho}
\delta^0_{\sigma}$ &\cr
\omit&height2pt&\omit&&\omit&&\omit&&\omit&\cr \tablerule
\omit&height2pt&\omit&&\omit&&\omit&&\omit&\cr && 5 && $H^4
g_{\mu\nu} g_{\rho\sigma}$ && 12 && $0$ &\cr
\omit&height2pt&\omit&&\omit&&\omit&&\omit&\cr \tablerule
\omit&height2pt&\omit&&\omit&&\omit&&\omit&\cr && 6 && $0$ && 13 &&
$H^4 [a^2 \delta^0_{\mu} \delta^0_{\nu} g_{\rho\sigma} + g_{\mu\nu}
a^2 \delta^0_{\rho} \delta^0_{\sigma}]$ &\cr
\omit&height2pt&\omit&&\omit&&\omit&&\omit&\cr \tablerule
\omit&height2pt&\omit&&\omit&&\omit&&\omit&\cr && 7 && $-2 H^4 a^2
\delta^0_{(\mu} g_{\nu) (\rho} \delta^0_{\sigma)}$ && 14 && $H^4
[a^2 \delta^0_{\mu} \delta^0_{\nu} g_{\rho\sigma} - g_{\mu\nu} a^2
\delta^0_{\rho} \delta^0_{\sigma}]$ &\cr
\omit&height2pt&\omit&&\omit&&\omit&&\omit&\cr \tablerule}}

\caption{Coincidence limits of each basis tensor.}

\label{tenscoin}

\end{table}

It is now just a matter of taking a few simple limits and combining
results from previous sections. For example, Table \ref{tenscoin}
implies that only the first and fifth of the de Sitter invariant
tensors are nonzero at coincidence. We can therefore read off the de
Sitter invariant part of the spin zero contribution by first
consulting expressions (\ref{C01}) and (\ref{C05}) then expressions
(\ref{DeltaW}), (\ref{DeltaMW}) and (\ref{DeltaS0}),
\begin{eqnarray} \lefteqn{i \Bigl[\mbox{}_{\mu\nu} \Delta^{\rm
dS,0}_{\rho\sigma}\Bigr](x;x) = 4 H^4 g_{\mu (\rho} g_{\sigma) \nu}
\times \mathcal{C}^1_0(0) + H^4 g_{\mu\nu} g_{\rho\sigma} \times
\mathcal{C}^5_0(0) } \\
& & \hspace{0cm} = 4 H^4 g_{\mu (\rho} g_{\sigma) \nu} \times
2 S_0''(0) + H^4 g_{\mu\nu} g_{\rho\sigma} \nonumber \\
& & \hspace{1.5cm} \times \Biggl\{ -2 S_0'(0) + 4 S_0''(0) - \frac{8
MW'(0)}{(D \!-\! 1) H^2} - \frac{2 W(0)}{(D \!-\! 2) (D \!-\! 1)
H^4} \Biggr\} , \qquad \\
& & \hspace{0cm} = g_{\mu (\rho} g_{\sigma) \nu} \times
\frac{H^{D-2}}{(4\pi)^{\frac{D}2}} \Bigl( \frac{D\!-\!2}{D \!-\!
1}\Bigr) \Biggl\{ - (S_0)_2^b \Biggr\} \nonumber \\
& & \hspace{.3cm} + g_{\mu\nu} g_{\rho\sigma} \! \times \!
\frac{H^{D-2}}{(4\pi)^{\frac{D}2}} \Bigl( \frac{D \!-\! 2}{D \!-\!
1} \Bigr) \Biggl\{ \frac{(S_0)_1^b}{2} \!-\! \frac{(S_0)_2^b}{2}
\!+\! \frac{2 (MW)_1^b}{D \!-\! 2} \!+\! \frac{W_2 \!-\! 2
W_1}{(D\!-\!2)^2} \Biggr\} . \qquad \label{dS0coinc}
\end{eqnarray}

The de Sitter invariant part of the spin two contribution follows
similarly from expressions (\ref{C21}) and (\ref{C25}),
\begin{eqnarray} \lefteqn{i \Bigl[\mbox{}_{\mu\nu} \Delta^{\rm
dS,2}_{\rho\sigma}\Bigr](x;x) = 4 H^4 g_{\mu (\rho} g_{\sigma) \nu}
\times \mathcal{C}^1_2(0) + H^4 g_{\mu\nu} g_{\rho\sigma} \times
\mathcal{C}^5_2(0) } \\
& & \hspace{0cm} = 4 H^4 g_{\mu (\rho} g_{\sigma) \nu} \times
-\frac{(D\!-\!3)^2 D (D \!+\! 1)}{4 (D \!-\! 2) (D \!-\! 1)} \,
G''(0) \nonumber \\
& & \hspace{5cm} + H^4 g_{\mu\nu} g_{\rho\sigma} \times \frac{(D
\!-\! 3)^2 (D \!+\! 1)}{(D \!-\! 2) (D \!-\! 1)} \, G''(0) \; .
\qquad \label{intermed}
\end{eqnarray}
From the definition (\ref{Gdef}) of $G(y)$ in terms of $S_2(y)$, and
the expansion (\ref{DeltaS2}) of $S_2(y)$, we find,
\begin{eqnarray}
\lefteqn{G''(0) = -(D\!+\!4) (D \!+\!6) S_2''''(0) + 3 (D \!+\! 1)
(D \!+\! 4) S_2'''(0) - \frac32 D (D \!+\! 1) S_2''(0) \; ,} \nonumber \\
& & = 3 \Bigl( \frac{D \!-\! 2}{D \!-\! 3}\Bigr)^2 \frac{
H^{D-6}}{(4 \pi)^{\frac{D}2}} \Biggl\{ (D \!+\! 4) (D \!+\! 6)
(S_2)^b_4 \nonumber \\
& & \hspace{4.5cm} - 3 (D \!+\! 1) (D \!+\! 4) (S_2)^b_3 \!+\! 2 D
(D \!+\! 1) (S_2)^b_2 \Biggr\} . \qquad
\end{eqnarray}
The tensor structure in (\ref{intermed}) is dictated by
tracelessness so we can write the final result as,
\begin{eqnarray}
\lefteqn{i \Bigl[\mbox{}_{\mu\nu} \Delta^{\rm dS,2}_{\rho\sigma}
\Bigr](x;x) = \Biggl[-D g_{\mu (\rho} g_{\sigma) \nu} + g_{\mu\nu}
g_{\rho\sigma} \Biggr] \times \frac{H^{D-2}}{(4 \pi)^{\frac{D}2}}
\frac{3 (D \!-\! 2) (D \!+\! 1)}{D \!-\! 1} } \nonumber \\
& & \hspace{0cm} \times \Biggl\{ (D \!+\! 4) (D \!+\! 6) (S_2)^b_4
\!-\! 3 (D \!+\! 1) (D \!+\! 4) (S_2)^b_3 \!+\! 2 D (D \!+\! 1)
(S_2)^b_2 \Biggr\} . \qquad \label{dS2coinc}
\end{eqnarray}
The coefficients $(S_2)^b_n$ are given in (\ref{S2bn}).

The de Sitter invariant contributions (\ref{dS0coinc}) and
(\ref{dS2coinc}) to the coincidence limit are divergent constants
times products of the metric. In contrast, the de Sitter breaking
terms give time dependence. This is most evident in the de Sitter
breaking contribution from the spin zero part,
\begin{eqnarray}
\lefteqn{ i \Bigl[\mbox{}_{\mu\nu} \Delta^{\rm
br,0}_{\rho\sigma}\Bigr](x;x) = H^4 g_{\mu\nu} g_{\rho\sigma} \times
\delta \mathcal{C}_0^5(u,0,0) - 2 H^4 a^2 \delta^0_{(\mu} g_{\nu)
(\rho} \delta^0_{\sigma)} \times \delta \mathcal{C}_0^7(u,0,0) }
\nonumber \\
& & \hspace{-.7cm} + H^4 a^4 \delta^0_{\mu} \delta^0_{\nu}
\delta^0_{\rho} \delta^0_{\sigma}\!\! \times \!\! \delta
\mathcal{C}_0^{11}(u,0,0) \!+\! H^4 \Bigl[a^2 \delta^0_{\mu}
\delta^0_{\nu} g_{\rho\sigma} \!+\! g_{\mu\nu} a^2 \delta^0_{\rho}
\delta^0_{\sigma} \Bigr] \!\!\times \!\! \delta
\mathcal{C}_0^{13}(u,0,0) \; . \qquad \label{dSb0coinc}
\end{eqnarray}
For $k=5$, 11 and 13 the coincident coefficient functions are,
\begin{eqnarray}
\lefteqn{\delta \mathcal{C}_0^k(u,0,0) = \frac{k_M}{H^4} \Biggl\{
\Bigl[(u \!+\! C_M) A_1^k \!+\! B_1^k \Bigr] e^{(b_M - b_A) u} }
\nonumber \\
& & \hspace{3.5cm} + \Bigl[ (u \!+\! C_M) (A_3^k \!-\! 2 A_2^k)
\!+\! B_3^k \!-\! 2 B_2^k\Bigr] e^{(b_M - b_W) u} \Biggr\} . \qquad
\label{11c13}
\end{eqnarray}
And for $k=7$ we have,
\begin{equation}
\delta \mathcal{C}_0^7(u,0,0) = \frac{k_M}{H^4} \Bigl[ (u \!+\! C_M)
A_2^k + B_2^k\Bigr] e^{(b_M - b_W) u} \; . \label{7c}
\end{equation}
Recall the definition (\ref{bdefs}) of $b_A$, $b_W$ and $b_M$, and
the definition (\ref{kdefs}) of $k_M$. We define $C_M \equiv 2
\psi(b_M) + 2 \ln(2)$, and the constants $A^k_{1-3}$ and $B^k_{1-3}$
can be found on Tables~\ref{spin0AB1}-\ref{spin0AB2}. It is worth
noting that there is no possibility of cancelations: de Sitter
breaking in the graviton propagator is a real and inevitable
phenomenon.

The de Sitter breaking contribution from the spin two part
(\ref{nondS2}) was much harder to derive, but its coincidence limit
is vastly simpler. Note first the coincidence limits of the tangent
space tensors (\ref{gperp}) and (\ref{Rperp}),
\begin{eqnarray}
\lim_{z \rightarrow x} g^{\perp}_{\rho\sigma}(z) & = &
g^{\perp}_{\mu\nu}(x) \equiv g_{\rho\sigma}(x) + a_x^2
\delta^0_{\rho} \delta^0_{\sigma} \; , \label{lim1} \\
\lim_{z \rightarrow x} \mathcal{R}^{\perp}_{\mu\nu}(x;z) & = &
g^{\perp}_{\mu\rho}(x) \; . \label{lim2}
\end{eqnarray}
Combining (\ref{nondS2}) with (\ref{lim1}-\ref{lim2}) implies,
\begin{equation}
i\Bigl[\mbox{}_{\mu\nu} \Delta^{\rm br,2}_{\rho\sigma}\Bigr](x;x) =
k \Biggl[ 2 \ln(2 a) \!+\! 2 \psi\Bigl( \frac{D \!-\!1}2\Bigr) \!-\!
4 \!+\! \frac1{D \!-\! 1}\Biggr] \Biggl\{ 2 g^{\perp}_{\mu (\rho}
g^{\perp}_{\sigma) \nu} \!-\! \frac{2 g^{\perp}_{\mu\nu}
g^{\perp}_{\rho\sigma}}{D \!-\! 1} \Biggr\} . \label{dSb2coinc}
\end{equation}

\section{Discussion}

The goal of this paper has been to facilitate dimensional
regularization computations of graviton loop diagrams on de Sitter
background using the de Donder gauge propagator of \cite{MTW3}. The
form of that propagator was summarized in section 2.4. Recall that
it consists of a spin zero part (\ref{Spin0}) and a spin two part
(\ref{Spin2}). Each part consists of a differential projector at
each coordinate --- these projectors are given in expressions
(\ref{spin0op}) and (\ref{spin2op}), respectively --- acting on the
appropriate structure function. Each structure function has a de
Sitter invariant and a de Sitter breaking part. Explicit series
expansions for the de Sitter invariant parts were derived in
expressions (\ref{DeltaS0}) and (\ref{DeltaS2}), respectively.
Explicit expressions for the de Sitter breaking parts of each
structure function were derived in expressions (\ref{deltaS0}) and
(\ref{deltaS2}), respectively.

To obtain a completely explicit expression for the graviton
propagator one must act the differential projectors on each
structure function. In our view, this step is unlikely to be
necessary in most real computations because there will be great
simplifications when free indices are contracted. We have therefore
contented ourselves with representing the de Sitter invariant
contributions as linear combinations of the five basis tensors in
Table~\ref{dStens} times coefficients which are expressed in terms
of derivatives of the de Sitter invariant parts of the structure
functions. These coefficients are expressions (\ref{C01}-\ref{C05})
and (\ref{C21}-\ref{C25}), respectively.

For the de Sitter breaking terms we derived completely explicit
expressions to emphasize that acting the differential projectors
does not annihilate them. The tensor structure consists of the five
de Sitter invariant basis tensors of Table~\ref{dStens} plus the
nine de Sitter breaking basis tensors of Table~\ref{nondStens}. The
coefficient functions for the spin zero are quite complicated,
taking the form described in expressions
(\ref{masterform}-\ref{submaster2}), with the coefficients given in
Table~\ref{spin0AB1}-\ref{spin0AB2}. The de Sitter breaking
contributions to the spin two part were more difficult to derive but
vastly simpler to state: they are given by equation (\ref{nondS2}).
The infrared logarithm on the mixed index parts of this expression
agrees precisely with the result obtained long ago, in a
noncovariant, average gauge \cite{RPW,TW2}. The same is likely to
hold in any gauge for which the cosmological symmetries of
homogeneity and isotropy are preserved.

The de Sitter invariant contributions to the coincidence limit of
the graviton propagator are given by expressions (\ref{dS0coinc}) and
(\ref{dS2coinc}). It consists of divergent constants times products
of the de Sitter metric. In contrast, both de Sitter breaking
contributions show time dependence. The contribution from the spin
zero part is given by relation (\ref{dSb0coinc}). The powers of the
scale factor which are evident in expressions (\ref{11c13}) and
(\ref{7c}) have no analogue in the one other gauge for which a
reliable solution for the complete propagator exists \cite{RPW,TW2},
so they can be regarded as peculiarities of de Donder gauge.
However, the contribution (\ref{dSb2coinc}) from the spin two part
shows exactly the same sorts of infrared logarithms as in the other
gauge. The appearance of these infrared logarithms in two vastly
different gauges supports the view that they are a gauge independent
feature of the theory.

The fact that there is a de Sitter breaking contribution even to the
spin two part from the propagator seems to contradict the recent claim
by Higuchi, Marolf and Morrison \cite{HMM} that free, dynamical
gravitons in synchronous-transverse-traceless gauge are physically
de Sitter invariant. Because our de Donder gauge condition is de
Sitter invariant, it cannot give rise to any compensating gauge
transformation which could cancel the explicit de Sitter breaking we
have exhibited. One might worry about de Sitter breaking through the
surface gauge conditions which are implicitly imposed in any covariant
gauge. However, we have followed the standard practice of simply
extending the $i \epsilon$ prescription to the gauge sector
\cite{TW6}, and it is difficult to see how that can introduce de
Sitter breaking if none was physically present.

We believe the more likely resolution of the disagreement is that
two physically different theories are being compared. Canonical
quantization of free gravitons in synchronous-transverse-traceless
gauge does show de Sitter breaking through the standard infrared
divergence of Bunch-Davies vacuum. Higuchi, Marolf and Morrison
avoid this by changing what they call ``the graviton field'' through
a nonlocal field redefinition they were rightly careful not to identify
as a linearized gauge transformation. Their new field obeys the same
equations of motion as the old one but it has different commutation
relations \cite{MTW4}. Of course adopting a noncanonical quantization
procedure results in a new theory, which is potentially physically
different from the original one. One of these differences is that the
``propagator'' between $x^{\mu}$ and $z^{\mu}$ obeys a sort of
dipole equation with a delta function source at $x^{\mu} = z^{\mu}$
and compensating anti-sources at temporal infinity \cite{MTW4}.
Adding anti-sources to the propagator equation will indeed result in
better infrared behavior, but it doesn't seem to be right physically.

It has long been obvious that de Sitter invariant gauges make
propagators vastly more complicated than gauges which exploit the
conformal flatness of de Sitter \cite{RPW,TW2}. It is easy to
quantify this observation by comparing the de Donder gauge result
\cite{MTW3} we have studied here with the de Sitter breaking gauge
\cite{TW2} in which all previous loop computations
\cite{TW3,TW4,MW1,KW2} have been done. Both propagators can be
expressed as linear combinations of basis tensors times coefficient
functions. They differ in three ways:
\begin{itemize}
\item{Only three basis tensors are needed for the de Sitter breaking
gauge, whereas 14 basis tensors are required for the invariant
gauge;}
\item{The basis tensors of the de Sitter breaking gauge are
constants, whereas those of the invariant gauge are complicated
functions of space and time; and}
\item{The three scalar coefficient functions of the de Sitter breaking
gauge are just $i\Delta_b(x;z)$ for $b_A = (\frac{D-1}2)$, $b_B=
(\frac{D-3}2)$ and $b_C = (\frac{D-5}2)$, whereas the 14 coefficient
functions of the invariant gauge are complicated linear combinations
of up to two derivatives of $i\Delta_b(x;z)$ with respect to $b$,
evaluated at $b_A$, $b_B$, $b_W = (\frac{D+1}2)$ and $b_M = \frac12
\sqrt{(D-1)(D+7)}$.}
\end{itemize}
To appreciate the final point, note that the infinite series
contributions in (\ref{expansion}) drop out in $D=4$ for
$i\Delta_A(x;z)$, $i\Delta_B(x;z)$ and $i\Delta_C(x;z)$. In fact
$i\Delta_A(x;z)$ consists of only two terms in $D=4$, while
$i\Delta_B(x;z)$ agrees with $i\Delta_C(x;z)$ for $D=4$, and they
have only a single term. However, differentiating with respect to
$b$ causes the infinite series to contribute, as it does for $b_M$,
even without derivatives. So the coefficient functions of the
invariant gauge seem horrifically more complex than those of the de
Sitter breaking gauge.

It is interesting to note that even the de Sitter breaking parts of
the invariant gauge are considerably more complicated than the
single factor of $\ln(a_x a_z)$ which occurs in $i\Delta_A(x;z)$.
However, not everything is in favor of the de Sitter breaking gauge.
One advantage of the de Donder gauge is that it admits only
invariant counterterms. It might also be that, when all the
derivatives are acted and all the contractions are performed in an
explicit computation involving propagators and vertices, the
complicated tensor factors drop out and the horrific structure
functions get acted upon by exactly the right differential operators
to produce simple results. That sounds like wishful thinking but
exactly these simplifications do occur when using the Lorentz gauge
photon propagator \cite{TW5} in one and two loop computations in
scalar QED \cite{PTW}.

\centerline{\bf Acknowledgements}

This work was partially supported by Marie Curie Grant IRG-247803,
by DFG-Research Training Group "Quantum and Gravitational Fields"
GRK 1523, by NWO Veni Project \# 680-47-406, by NSF grant PHY-0855021, 
and by the Institute for Fundamental Theory at the University of Florida.

\section{Appendix}

We here give a number of relations that are used in sections 4 and 5
to act the transverse-traceless projectors
$\mathbf{P}_{\mu\nu}^{~~\alpha\beta}(x) \times
\mathbf{P}_{\rho\sigma}^{~~\kappa\lambda}(z)$ on the spin two
structure function. The first set of 12 equalities follow from the
differentiation and contraction identities of subsection 2.1, and
hold for any function $F$, whether or not it is de Sitter invariant.
These 12 relations, which are assumed to be symmetrized on $\mu$ and
$\nu$ where both indices appear, were employed in the derivation of
equation (\ref{1stP}),
\begin{eqnarray}
\lefteqn{ \frac{D}{D x^{\alpha}} \Bigl[ \Bigl(
\mathcal{R}^{\alpha}_{~\kappa} \mathcal{R}_{\nu \lambda} \!+\!
\mathcal{R}_{\nu \kappa} \mathcal{R}^{\alpha}_{~\lambda} \Bigr)
F\Bigr] = (D \!+\! 1) \mathcal{R}_{\nu (\kappa} \frac{\partial
y}{\partial z^{\lambda)}} F + 2 \mathcal{R}_{\nu (\kappa}
\mathcal{R}^{\alpha}_{~\lambda)} \frac{D F}{Dx^{\alpha}} \; , }
\label{ID1} \\
\lefteqn{\frac{D}{Dx^{\mu}} \Bigl[ \mathcal{R}_{\nu (\kappa}
\frac{\partial y}{\partial z^{\lambda)}} F\Bigr] = \frac{
g_{\mu\nu}}{2} \frac{\partial y}{\partial z^{\kappa}} \frac{\partial
y}{\partial z^{\lambda}} F \!-\! 2H^2 \mathcal{R}_{\mu\kappa}
\mathcal{R}_{\nu\lambda} F \!+\! \mathcal{R}_{\mu (\kappa}
\frac{\partial y}{\partial z^{\lambda)}} \frac{DF}{Dx^{\nu}} \; ,
\qquad \label{ID2} } \\
\lefteqn{ \frac{D}{Dx^{\mu}} \Bigl[ \mathcal{R}_{\nu (\kappa}
\mathcal{R}^{\alpha}_{~ \lambda)} F\Bigr] = \frac12 g_{\mu\nu}(x)
\mathcal{R}^{\alpha}_{~ (\kappa} \frac{\partial y}{\partial
z^{\lambda)}} F + \frac12 \delta^{\alpha}_{~(\mu} \mathcal{R}_{\nu)
(\kappa} \frac{\partial y}{\partial z^{\lambda)}} F} \nonumber \\
& & \hspace{8.5cm} + \mathcal{R}_{\mu (\kappa}
\mathcal{R}^{\alpha}_{~\lambda)} \frac{DF}{D x^{\nu}} \; , \qquad
\label{ID3} \\
\lefteqn{ \frac{D}{D x^{\mu}} \Bigl[ \frac{\partial y}{\partial
z^{\kappa}} \frac{\partial y}{\partial z^{\lambda}} F\Bigr] = -4 H^2
\mathcal{R}_{\mu (\kappa} \frac{\partial y}{\partial z^{\lambda)}} F
+ \frac{\partial y}{\partial z^{\kappa}} \frac{\partial y}{\partial
z^{\lambda}} \frac{DF}{Dx^{\mu}} \; , \qquad \label{ID4}} \\
\lefteqn{ \frac{D^2}{Dx^{\alpha} Dx^{\beta}} \Bigl[
\mathcal{R}^{\alpha}_{~\kappa} \mathcal{R}^{\beta}_{~\lambda}
F\Bigr] = -(D \!+\! 1) H^2 g_{\kappa\lambda}(z) F + \Bigl(\frac{D
\!+\! 1}{2}\Bigr)^2 \frac{\partial y}{\partial z^{\kappa}}
\frac{\partial y}{\partial z^{\lambda}} F } \nonumber \\
& & \hspace{4cm} + (D \!+\! 1) \mathcal{R}^{\alpha}_{~ (\kappa}
\frac{\partial y}{\partial z^{\lambda)}} \frac{DF}{Dx^{\alpha}} +
\mathcal{R}^{\alpha}_{~\kappa} \mathcal{R}^{\beta}_{~\lambda}
\frac{D^2 F}{Dx^{\alpha} Dx^{\beta}} \; , \qquad \label{ID5} \\
\lefteqn{\square_x \Bigl[ \mathcal{R}_{\mu \kappa} \mathcal{R}_{\nu
\lambda} F \Bigr] = \frac12 g_{\mu\nu}(x) \frac{\partial y}{\partial
z^{\kappa}} \frac{\partial y}{\partial z^{\lambda}} F - 2 H^2
\mathcal{R}_{\mu \kappa} \mathcal{R}_{\nu \lambda} F } \nonumber \\
& & \hspace{3.5cm} + \mathcal{R}_{\mu\kappa} \frac{\partial
y}{\partial z^{\lambda}} \frac{DF}{Dx^{\nu}} + \mathcal{R}_{\nu
\lambda} \frac{\partial y}{\partial z^{\kappa}} \frac{DF}{Dx^{\mu}}
+ \mathcal{R}_{\mu\kappa} \mathcal{R}_{\nu\lambda} \square_x F \; ,
\qquad \label{ID6} \\
\lefteqn{\square_x \Bigl[ \mathcal{R}_{\nu (\kappa} \frac{\partial
y}{\partial z^{\lambda)}} F \Bigr] = -(D\!+\!3) H^2 \mathcal{R}_{\nu
(\kappa} \frac{\partial y}{\partial z^{\lambda)}} F } \nonumber \\
& & \hspace{2cm} + \frac{\partial y}{\partial z^{\kappa}}
\frac{\partial y}{\partial z^{\lambda}} \frac{DF}{Dx^{\nu}} - 4H^2
\mathcal{R}_{\nu (\kappa} \mathcal{R}^{\alpha}_{~\lambda)}
\frac{DF}{Dx^{\alpha}} + \mathcal{R}_{\nu (\kappa} \frac{\partial
y}{\partial z^{\lambda)}} \square_x F \; , \qquad \label{ID7} \\
\lefteqn{\square_x \Bigl[ \frac{\partial y}{\partial z^{\kappa}}
\frac{\partial y}{\partial z^{\lambda}} F\Bigr] = 8 H^4
g_{\kappa\lambda}(z) F - 2 (D \!+\! 1) H^2 \frac{\partial
y}{\partial z^{\kappa}} \frac{\partial y}{\partial
z^{\lambda}} F } \nonumber \\
& & \hspace{5cm} - 8 H^2 \mathcal{R}^{\alpha}_{~ (\kappa}
\frac{\partial y}{\partial z^{\lambda)}} \frac{DF}{Dx^{\alpha}} +
\frac{\partial y}{\partial z^{\kappa}} \frac{\partial
y}{\partial z^{\lambda}} \square_x F \; , \qquad \label{ID8} \\
\lefteqn{ \frac{D^2}{D x^{\mu} Dx^{\nu}} \Bigl[ \frac{\partial
y}{\partial z^{\kappa}} \frac{\partial y}{\partial z^{\lambda}} F
\Bigr] = -2 H^2 g_{\mu\nu}(x) \frac{\partial y}{\partial z^{\kappa}}
\frac{\partial y}{\partial z^{\lambda}} F + 8 H^4 \mathcal{R}_{\mu
\kappa} \mathcal{R}_{\nu \lambda} F } \nonumber \\
& & \hspace{4.5cm} -8 H^2 \mathcal{R}_{\mu (\kappa} \frac{\partial
y}{\partial z^{\lambda)}} \frac{DF}{Dx^{\nu}} + \frac{\partial
y}{\partial z^{\kappa}} \frac{\partial y}{\partial z^{\lambda}}
\frac{D^2 F}{Dx^{\mu} Dx^{\nu}} \; , \qquad \label{ID9} \\
\lefteqn{ \frac{D^2}{Dx^{\mu} Dx^{\nu}} \Bigl[
\mathcal{R}^{\alpha}_{~ (\kappa} \frac{\partial y}{\partial
z^{\lambda)}} F \Bigr] = - H^2 g_{\mu\nu}(x)
\mathcal{R}^{\alpha}_{~(\kappa} \frac{\partial y}{\partial
z^{\lambda)}} F - 3 H^2 \delta^{\alpha}_{~(\mu} \mathcal{R}_{\nu)
(\kappa} \frac{\partial y}{\partial z^{\lambda)}} F} \nonumber \\
& & \hspace{.5cm} + \frac{\partial y}{\partial z^{\kappa}}
\frac{\partial y}{\partial z^{\lambda}} \delta^{\alpha}_{~(\mu}
\frac{DF}{Dx^{\nu)}} - 4 H^2 \mathcal{R}_{\mu (\kappa}
\mathcal{R}^{\alpha}_{~\lambda)} \frac{DF}{D x^{\nu}} +
\mathcal{R}^{\alpha}_{~ (\kappa} \frac{\partial y}{\partial
z^{\lambda)}} \frac{D^2 F}{Dx^{\mu} Dx^{\nu}} \; . \qquad
\label{ID10} \\
\lefteqn{ \square_x \Bigl[ \mathcal{R}^{\alpha}_{~(\kappa}
\mathcal{R}^{\beta}_{~\lambda)} F\Bigr] = -2 H^2
\mathcal{R}^{\alpha}_{~(\kappa} \mathcal{R}^{\beta}_{~\lambda)} F +
\frac12 g^{\alpha\beta}(x) \frac{\partial y}{\partial z^{\kappa}}
\frac{\partial y}{\partial z^{\lambda}} F } \nonumber \\
& & \hspace{5.2cm} + 2 \frac{\partial y}{\partial z^{(\kappa}}
\mathcal{R}^{(\alpha}_{~~\lambda)} D^{\beta)}_x F +
\mathcal{R}^{\alpha}_{~(\kappa} \mathcal{R}^{\beta}_{~\lambda)}
\square_x F \; , \qquad \label{ID11} \\
\lefteqn{ \frac{D^2}{Dx^{\mu} Dx^{\nu}} \Bigl[
\mathcal{R}^{\alpha}_{~(\kappa} \mathcal{R}^{\beta}_{~\lambda)}
F\Bigr] = -2 H^2 \delta^{(\alpha}_{~~(\mu} \mathcal{R}_{\nu)
(\kappa} \mathcal{R}^{\beta)}_{~\lambda)} F + \frac12
\delta^{\alpha}_{~(\mu} \delta^{\beta}_{~\nu)} \frac{\partial
y}{\partial z^{\kappa}} \frac{\partial y}{\partial z^{\lambda}} F}
\nonumber \\
& & \hspace{3.5cm} + 2 \frac{\partial y}{\partial z^{(\kappa}}
\mathcal{R}^{(\alpha}_{~~\lambda)} \delta^{\beta)}_{~~(\mu}
\frac{DF}{Dx^{\nu)}} + \mathcal{R}^{\alpha}_{~(\kappa}
\mathcal{R}^{\beta}_{~\lambda)} \frac{D^2 F}{Dx^{(\mu} Dx^{\nu)}} \;
. \qquad \label{ID12}
\end{eqnarray}

Eight additional simplifications follow from two assumptions:
\begin{itemize}
\item{That the result is being acted upon by $\mathbf{P}_{\rho\sigma}^{
\kappa\lambda}(z)$, so that we can neglect factors of $g_{\kappa\lambda}(z)$
and total derivatives $D/Dz^{\kappa}$ or $D/Dz^{\lambda}$; and}
\item{The function $F$ depends only upon the de Sitter invariant $y(x;z)$.}
\end{itemize}
To save space we omit the factors of $\mathbf{P}_{~~\rho\sigma}^{
\kappa\lambda}(z)$ which act on the right and left, and we symmetrize on
the indices $\mu$ and $\nu$ where both appear,
\begin{eqnarray}
\frac{\partial y}{\partial z^{\kappa}} \frac{\partial y}{\partial z^{\lambda}}
& \Longrightarrow & - 4 H^2 \mathcal{R}^{\alpha}_{~\kappa}
\mathcal{R}_{\alpha \lambda} \; , \qquad \label{SID1} \\
\mathcal{R}_{\mu \kappa} \frac{\partial y}{\partial z^{\lambda}}
\frac{D F}{D x^{\nu}} & \Longrightarrow & 2 H^2 \mathcal{R}_{\mu\kappa}
\mathcal{R}_{\nu \lambda} F \; , \qquad \label{SID2} \\
\mathcal{R}^{\alpha}_{~\kappa} \frac{\partial y}{\partial z^{\lambda}}
\frac{D^3 F}{Dx^{\mu} Dx^{\nu} Dx^{\alpha}} & \Longrightarrow &
2 H^2 \mathcal{R}^{\alpha}_{~\kappa} \mathcal{R}_{\alpha\lambda} \Bigl[
\frac{D^2 F}{D x^{\mu} D x^{\nu}} + H^2 g_{\mu\nu} F \Bigr] \nonumber \\
& & \hspace{-1cm} + 4 H^2 \mathcal{R}^{\alpha}_{~ \kappa}
\mathcal{R}_{\mu \lambda} \frac{D^2 F}{Dx^{\nu} Dx^{\alpha}}
+ 2 H^4 \mathcal{R}_{\mu\kappa} \mathcal{R}_{\nu \lambda} F \; , \qquad
\label{SID3} \\
\mathcal{R}^{\alpha}_{~\kappa} \mathcal{R}_{\alpha \lambda}
\frac{D^2 F}{D x^{\mu} D x^{\nu}} & \Longrightarrow &
-2 H^2 \mathcal{R}_{\mu\kappa} \mathcal{R}_{\nu \lambda} F + H^2 g_{\mu\nu}
\mathcal{R}^{\alpha}_{~\kappa} \mathcal{R}_{\alpha\lambda} F \; , \qquad
\label{SID4} \\
\mathcal{R}^{\alpha}_{~\kappa} \mathcal{R}_{\alpha \lambda} \square F
& \Longrightarrow & (D \!-\! 2) H^2 \mathcal{R}^{\alpha}_{~\kappa}
\mathcal{R}_{\alpha\lambda} F \; , \qquad \label{SID5} \\
\mathcal{R}_{\mu\kappa} \mathcal{R}^{\alpha}_{~\lambda}
\frac{D^2 F}{Dx^{\nu} Dx^{\alpha}} & \Longrightarrow & - H^2
\mathcal{R}_{\mu\kappa} \mathcal{R}_{\nu\lambda} F \; , \qquad \label{SID6} \\
\mathcal{R}^{\alpha}_{~\kappa} \mathcal{R}^{\beta}_{~\lambda}
\frac{D^2 F}{Dx^{\alpha} Dx^{\beta}} & \Longrightarrow & - H^2
\mathcal{R}^{\alpha}_{~\kappa} \mathcal{R}_{\alpha\lambda} F \; ,
\qquad \label{SID7} \\
\mathcal{R}^{\alpha}_{~\kappa} \mathcal{R}^{\beta}_{~\lambda}
\frac{D^4 F}{Dx^{\alpha} Dx^{\beta} Dx^{\mu} D x^{\nu} } & \Longrightarrow &
2 H^4 \mathcal{R}_{\mu\kappa} \mathcal{R}_{\nu\lambda} F - H^4 g_{\mu\nu}
\mathcal{R}^{\alpha}_{~\kappa} \mathcal{R}_{\alpha\lambda} F \; .
\qquad \label{SID8}
\end{eqnarray}
These identities were used to derive equation (\ref{1stPsimp}).

When the structure function depends on $u$, rather than $y$, we
cannot make (\ref{1stPsimp}). The result we require for that case is
equation (\ref{newP2}). It can be derived using the following
identities,
\begin{eqnarray}
\lefteqn{\frac{D}{D z^{\rho}} \Bigl[ \mathcal{R}_{\mu \kappa}
\frac{\partial y}{\partial z^{\lambda}} G\Bigr] = \mathcal{R}_{\mu
\kappa} \frac{\partial y}{\partial z^{\lambda}} \frac{D G}{D
z^{\rho}} \!+\! \frac{g_{\rho\kappa}}{2} \frac{\partial y}{\partial
x^{\mu}} \frac{\partial y}{\partial z^{\lambda}} G \!+\! H^2 (2
\!-\! y) g_{\rho \lambda} \mathcal{R}_{\mu \kappa} G \; , \qquad}
\label{newID1} \\
\lefteqn{\frac{D}{D z^{\rho}} \Bigl[ \frac{\partial y}{\partial
x^{\mu}} \frac{\partial y}{\partial z^{\kappa}} G \Bigr] =
\frac{\partial y}{\partial x^{\mu}} \frac{\partial y}{\partial
z^{\kappa}} \frac{D G}{D z^{\rho}} \!-\! 2 H^2 \mathcal{R}_{\mu
\rho} \frac{\partial y}{\partial z^{\kappa}} G \!+\! H^2 (2 \!-\! y)
g_{\rho \kappa} \frac{\partial y}{\partial x^{\mu}} G \; , }
\label{newID2} \\
\lefteqn{\frac{D}{D z^{\rho}} \Bigl[ (2 \!-\! y)
\mathcal{R}_{\mu\kappa} G \Bigr] = (2 \!-\! y)
\mathcal{R}_{\mu\kappa} \frac{D G}{D z^{\rho}} \!-\!
\mathcal{R}_{\mu\kappa} \frac{\partial y}{\partial z^{\rho}} G \!+\!
\frac12 (2 \!-\! y) g_{\rho\kappa} \frac{\partial y}{\partial
x^{\mu}} G \; , } \label{newID3} \\
\lefteqn{\frac{D}{D z^{\rho}} \Bigl[ (2 \!-\! y) \frac{\partial
y}{\partial x^{\mu}} G \Bigr] = (2 \!-\! y) \frac{\partial
y}{\partial x^{\mu}} \frac{DG}{D z^{\rho}} \!-\! \frac{\partial
y}{\partial x^{\mu}} \frac{\partial y}{\partial z^{\rho}} G \!-\! 2
H^2 (2 \!-\! y) \mathcal{R}_{\mu\rho} G \; , }
\label{newID4} \\
\lefteqn{\square_z \Bigl[ \mathcal{R}_{\mu \kappa} \frac{\partial
y}{\partial z^{\lambda}} G\Bigr] = \mathcal{R}_{\mu \kappa}
\frac{\partial y}{\partial z^{\lambda}} \Bigl[\square_z \!-\! 2 H^2
\Bigr] G + \frac{\partial y}{\partial x^{\mu}} \frac{\partial
y}{\partial z^{\kappa}} \frac{D G}{D z^{\lambda}} } \nonumber \\
& & \hspace{4cm} + 2 H^2 (2 \!-\! y) \mathcal{R}_{\mu \kappa}
\frac{D G}{D z^{\lambda}} + H^2 (2 \!-\! y) g_{\kappa\lambda}
\frac{\partial y}{\partial x^{\mu}} G \; , \qquad \label{newID5} \\
\lefteqn{\square_z \Bigl[ \frac{\partial y}{\partial x^{\mu}}
\frac{\partial y}{\partial z^{\kappa}} G \Bigr] = \frac{\partial y}{
\partial x^{\mu}} \frac{\partial y}{\partial z^{\kappa}} \Bigl[
\square_z \!-\! (D \!+\! 1) H^2\Bigr] G \!-\! 4 H^2
\mathcal{R}_{\mu}^{~\lambda} \frac{\partial y}{\partial z^{\kappa}}
\frac{D G}{D z^{\lambda}} } \nonumber \\
& & \hspace{4cm} + 2 H^2 (2 \!-\! y) \frac{\partial y}{\partial
x^{\mu}} \frac{D G}{D z^{\kappa}} \!-\! 4 H^4 (2 \!-\! y)
\mathcal{R}_{\mu\kappa} G\; , \qquad \label{newID6} \\
\lefteqn{\square_z \Bigl[ (2 \!-\! y) \mathcal{R}_{\mu\kappa} G
\Bigr] = (2 \!-\! y) \mathcal{R}_{\mu\kappa} \Bigl[ \square_z \!-\!
(D \!+\! 1) H^2\Bigr] G \!-\! 2 \mathcal{R}_{\mu\kappa}
\frac{\partial y}{\partial z^{\lambda}} \frac{D G}{D z_{\lambda}} }
\nonumber \\
& & \hspace{6cm} + (2 \!-\! y) \frac{\partial y}{\partial x^{\mu}}
\frac{D G}{D z^{\kappa}} \!-\! \frac{\partial y}{\partial x^{\mu}}
\frac{\partial y}{\partial z^{\kappa}} G \; , \qquad
\label{newID7} \\
\lefteqn{\square_z \Bigl[ (2 \!-\! y) \frac{\partial y}{\partial
x^{\mu}} G \Bigr] = (2 \!-\! y) \frac{\partial y}{\partial x^{\mu}}
\Bigl[ \square_z \!-\! 2 H^2\Bigr] G \!-\! 2 \frac{\partial
y}{\partial x^{\mu}} \frac{\partial y}{\partial z^{\lambda}} \frac{D
G}{D z_{\lambda}} } \nonumber \\
& & \hspace{4cm} - 4 H^2 (2 \!-\! y) \mathcal{R}_{\mu}^{~\lambda}
\frac{D G}{D z^{\lambda}} \!-\! 2 D H^2 (2 \!-\! y) \frac{\partial
y}{\partial x^{\mu}} G \; . \qquad \label{newID8}
\end{eqnarray}
Note that tracing on $\kappa$ and $\lambda$ gives relation
(\ref{newID4}) from (\ref{newID1}) and relation (\ref{newID8}) from
(\ref{newID5}).

It remains to act the derivatives. We can save some space by
defining the transverse metric,
\begin{equation}
g^{\perp}_{\alpha\beta}(x) \equiv g_{\alpha\beta}(x) + \frac1{H^2}
\frac{\partial u}{\partial x^{\alpha}} \frac{\partial u}{\partial
x^{\beta}} \quad , \quad g^{\perp}_{\kappa\lambda}(z) \equiv
g_{\kappa\lambda}(z) + \frac1{H^2} \frac{\partial u}{\partial
z^{\kappa}} \frac{\partial u}{\partial z^{\lambda}} \; .
\end{equation}
Another space-saving convention is to assume the $x$ indices are
$\alpha$, $\beta$, $\gamma$ and $\delta$, and that the $z$ indices
are $\kappa$, $\lambda$, $\theta$ and $\phi$. This allows us to
dispense with coordinates, for example,
\begin{equation}
g^{\perp}_{\alpha\beta}(x) \longrightarrow g^{\perp}_{\alpha\beta}
\qquad , \quad \frac{\partial u}{\partial x^{\alpha}}
\longrightarrow u_{\alpha} \qquad , \qquad \frac{\partial
u}{\partial z^{\kappa}} \longrightarrow u_{\kappa} \; .
\end{equation}
We can also take advantage of the fact that $F(u)$ is a third order
polynomial in $u$, so one gets zero for $F''''$ and all higher
derivatives. Finally, note that every term contains at least one
derivative with respect to $x$ and another with respect to $z$. The
ten cases we require are therefore,
\begin{eqnarray}
\lefteqn{D^x_{\alpha} D^z_{\kappa} F = u_{\alpha} u_{\kappa} F'' \; , } \\
\lefteqn{D^x_{\alpha} D^x_{\beta} D^z_{\kappa} F = -H^2
g^{\perp}_{\alpha\beta} u_{\kappa} F'' + u_{\alpha} u_{\beta}
u_{\kappa} F''' \; , } \\
\lefteqn{D^x_{\alpha} D^x_{\beta} D^x_{\gamma} D^z_{\kappa} F = 2
H^2 g^{\perp}_{\alpha (\beta} u_{\gamma)} u_{\kappa} F'' - 3 H^2
g^{\perp}_{(\alpha \beta} u_{\gamma)} u_{\kappa} F'''} \\
\lefteqn{D^x_{\alpha} D^x_{\beta} D^x_{\gamma} D^x_{\delta}
D^z_{\kappa} F = -2 \Bigl[H^4 g^{\perp}_{\alpha (\gamma}
g^{\perp}_{\delta) \beta} \!+\! H^2 g^{\perp}_{\alpha (\gamma}
u_{\delta)} u_{\beta} \!+\! H^2 g^{\perp}_{\alpha \beta} u_{\gamma}
u_{\delta} \Bigr] u_{\kappa} F'' } \nonumber \\
& & \hspace{2cm} + \Bigl[ 3 H^4 g^{\perp}_{(\alpha \beta}
g^{\perp}_{\gamma\delta)} \!+\! 6 H^2 g^{\perp}_{\alpha (\beta}
u_{\gamma} u_{\delta)} \!+\! 2 H^2 g^{\perp}_{\beta (\gamma}
u_{\delta)} u_{\alpha} \Bigr] u_{\kappa} F''' \; , \qquad \\
\lefteqn{D^x_{\alpha} D^x_{\beta} D^z_{\kappa} D^z_{\lambda} F = H^4
g^{\perp}_{\alpha\beta} g^{\perp}_{\kappa\lambda} F'' - \Bigl[ H^2
g^{\perp}_{\alpha\beta} u_{\kappa} u_{\lambda} \!+\! H^2
u_{\alpha} u_{\beta} g^{\perp}_{\kappa\lambda} \Bigr] F''' \; , } \\
\lefteqn{D^x_{\alpha} D^x_{\beta} D^x_{\gamma} D^z_{\kappa}
D^z_{\lambda} F = - 2 H^4 g^{\perp}_{\alpha (\beta} u_{\gamma)}
g^{\perp}_{\kappa\lambda} F''} \nonumber \\
& & \hspace{4.7cm} + \Bigl[ 3 H^4 g^{\perp}_{(\alpha\beta}
u_{\gamma)} g_{\kappa\lambda} \!+\! 2 H^2 g^{\perp}_{\alpha (\beta}
u_{\gamma)} u_{\kappa} u_{\lambda} \Bigr] F''' \; , \qquad \\
\lefteqn{D^x_{\alpha} D^x_{\beta} D^x_{\gamma} D^x_{\delta}
D^z_{\kappa} D^z_{\lambda} F = 2 \Bigl[H^6 g^{\perp}_{\alpha
(\gamma} g^{\perp}_{\delta) \beta} \!+\! H^4 g^{\perp}_{\alpha
(\gamma} u_{\delta)} u_{\beta} \!+\! H^4 g^{\perp}_{\alpha \beta}
u_{\gamma} u_{\delta} \Bigr] g^{\perp}_{\kappa\lambda} F ''} \nonumber \\
& & \hspace{0cm} - \Bigl[3 H^6 g^{\perp}_{\alpha (\beta}
g^{\perp}_{\gamma\delta)} g^{\perp}_{\kappa\lambda} \!+\! 6 H^4
g^{\perp}_{\alpha (\beta} u_{\gamma} u_{\delta)}
g^{\perp}_{\kappa\lambda} + 2 H^4 g^{\perp}_{\beta (\gamma}
u_{\delta)} u_{\alpha}
g^{\perp}_{\kappa\lambda} \nonumber \\
& & \hspace{.5cm} + 2 H^4 g^{\perp}_{\alpha (\gamma}
g^{\perp}_{\delta) \beta} u_{\kappa} u_{\lambda} \!+\! 2 H^2
g^{\perp}_{\alpha (\gamma} u_{\delta)} u_{\beta} u_{\kappa}
u_{\lambda} \!+\! 2 H^2 g^{\perp}_{\alpha\beta}
u_{\gamma} u_{\delta} u_{\kappa} u_{\lambda} \Bigr] F''' \; , \qquad \\
\lefteqn{D^x_{\alpha} D^x_{\beta} D^x_{\gamma} D^z_{\kappa}
D^z_{\lambda} D^z_{\theta} F = 4 H^4 g^{\perp}_{\alpha (\beta}
u_{\gamma)} g^{\perp}_{\kappa (\lambda} u_{\theta)} F''} \nonumber \\
& & \hspace{3.5cm} - \Bigl[6 H^4 g^{\perp}_{\alpha (\beta}
u_{\gamma)} g^{\perp}_{(\kappa \lambda} u_{\theta)} \!+\! 6 H^4
g^{\perp}_{(\alpha\beta}
u_{\gamma)} g^{\perp}_{\kappa (\lambda} u_{\theta)} \Bigr] F''' \; , \qquad \\
\lefteqn{D^x_{\alpha} D^x_{\beta} D^x_{\gamma} D^x_{\delta}
D^z_{\kappa} D^z_{\lambda} D^z_{\theta} F = -4 H^4 \Bigl[ H^2
g^{\perp}_{\alpha (\gamma} g^{\perp}_{\delta) \beta} \!+\!
g^{\perp}_{\alpha (\gamma} u_{\delta)} u_{\beta} \!+\!
g^{\perp}_{\alpha \beta} u_{\gamma} u_{\delta} \Bigr]
g^{\perp}_{\kappa (\lambda} u_{\theta)} F'' } \nonumber \\
& & \hspace{-.5cm} + \Bigl[6 H^6 g^{\perp}_{\alpha (\gamma}
g^{\perp}_{\delta) \beta} g^{\perp}_{(\kappa\lambda} u_{\theta)}
\!+\! 6 H^4 g^{\perp}_{\alpha (\gamma} u_{\delta)} u_{\beta}
g^{\perp}_{(\kappa \lambda} u_{\theta)} \!+\! 6 H^4
g^{\perp}_{\alpha\beta} u_{\gamma} u_{\delta}
g^{\perp}_{(\kappa \lambda} u_{\theta)} \nonumber \\
& & \hspace{-.5cm} + 6 H^6 g^{\perp}_{\alpha (\beta}
g^{\perp}_{\gamma\delta)} g^{\perp}_{\kappa (\lambda} u_{\theta)}
\!+\! 12 H^4 g^{\perp}_{\alpha (\beta} u_{\gamma} u_{\delta)}
g^{\perp}_{\kappa (\lambda} u_{\theta)} \!+\! 4 H^4 g^{\perp}_{\beta
(\gamma} u_{\delta)} u_{\alpha} g^{\perp}_{\kappa (\lambda}
u_{\theta)} \Bigr] F''' , \qquad \\
\lefteqn{D^x_{\alpha} D^x_{\beta} D^x_{\gamma} D^x_{\delta}
D^z_{\kappa} D^z_{\lambda} D^z_{\theta} D^z_{\phi} F = 4 H^4 \Bigl[
H^2 g^{\perp}_{\alpha (\gamma} g^{\perp}_{\delta) \beta} \!+\!
g^{\perp}_{\alpha (\gamma} u_{\delta)} u_{\beta} \!+\!
g^{\perp}_{\alpha\beta}
u_{\gamma} u_{\delta} \Bigr] } \nonumber \\
& & \hspace{0cm} \times \Bigl[ H^2 g^{\perp}_{\kappa (\theta}
g^{\perp}_{\phi) \lambda} \!+\! g^{\perp}_{\kappa (\theta} u_{\phi)}
u_{\lambda} \!+\! g^{\perp}_{\kappa\lambda} u_{\theta} u_{\phi}
\Bigr] F'' - \Bigl[6 H^8 g^{\perp}_{\alpha (\gamma}
g^{\perp}_{\delta) \beta}
g^{\perp}_{\kappa (\lambda} g^{\perp}_{\theta \phi)} \nonumber \\
& & \hspace{0cm} + 6 H^8 g^{\perp}_{\alpha (\beta} g^{\perp}_{\gamma
\delta)} g^{\perp}_{\kappa (\theta} g^{\perp}_{\phi) \lambda} \!+\!
4 H^6 g^{\perp}_{\alpha (\gamma} g^{\perp}_{\delta) \beta} \Bigl(
g^{\perp}_{\lambda (\theta} u_{\phi)} u_{\kappa} \!+\!
3 g^{\perp}_{\kappa (\lambda} u_{\theta} u_{\phi)} \Bigr) \nonumber \\
& & \hspace{0cm} + 6 H^6 \Bigl( g^{\perp}_{\alpha (\gamma}
u_{\delta)} u_{\beta} \!+\! g^{\perp}_{\alpha\beta} u_{\gamma}
u_{\delta}\Bigr) g^{\perp}_{\kappa (\lambda} g^{\perp}_{\theta
\phi)} \!+\! 6 H^6
g^{\perp}_{\alpha (\beta} g^{\perp}_{\gamma\delta)} \nonumber \\
& & \hspace{0cm} \times \Bigl( g^{\perp}_{\kappa (\theta} u_{\phi)}
u_{\lambda} \!+\! g^{\perp}_{\kappa\lambda} u_{\theta}
u_{\phi}\Bigr) + 4 H^6 \Bigl( 3 g^{\perp}_{\alpha (\beta} u_{\gamma}
u_{\delta)} \!+\! g_{\beta (\gamma} u_{\delta)} u_{\alpha} \Bigr)
g^{\perp}_{\kappa (\theta}
g^{\perp}_{\phi) \lambda} \nonumber \\
& & \hspace{0cm} + 4 H^4 \Bigl( g^{\perp}_{\alpha (\gamma}
u_{\delta)} u_{\beta} \!+\! g^{\perp}_{\alpha\beta} u_{\gamma}
u_{\delta}\Bigr) \Bigl( 3 g^{\perp}_{\kappa (\lambda} u_{\theta}
u_{\phi)} \!+\! g^{\perp}_{\lambda (\theta} u_{\phi)} u_{\kappa}
\Bigr)\nonumber \\
& & \hspace{2cm} + 4 H^4 \Bigl( 3 g^{\perp}_{\alpha (\beta}
u_{\gamma} u_{\delta)} \!+\! g^{\perp}_{\beta (\gamma} u_{\delta)}
u_{\alpha}\Bigr) \Bigl( g^{\perp}_{\kappa (\theta} u_{\phi)}
u_{\lambda} \!+\! g^{\perp}_{\kappa\lambda} u_{\theta} u_{\phi}
\Bigr) \Bigr] F''' \; . \qquad
\end{eqnarray}

\end{document}